\pdfoutput=1
\documentclass[letterpaper,journal]{IEEEtran}
\RequirePackage{siunitx}
\PassOptionsToPackage{detect-all, per-mode=symbol, range-units=single, range-phrase=~to~}{siunitx}
\RequirePackage{glossaries}
\PassOptionsToPackage{acronyms,nohypertypes={acronym},nomain}{glossaries}
\RequirePackage[pscoord]{eso-pic}
\RequirePackage{multirow}
\RequirePackage{makecell}
\RequirePackage{threeparttable}
\RequirePackage{booktabs}
\RequirePackage{colortbl}
\RequirePackage[verbose]{placeins}
\RequirePackage{fancyhdr}
\RequirePackage[hidelinks]{hyperref}
\RequirePackage{MnSymbol}
\RequirePackage{wasysym}
\RequirePackage{xspace}
\RequirePackage{lipsum}
\RequirePackage{soul}
\RequirePackage{calc}
\RequirePackage{float}
\RequirePackage{threeparttable}
\RequirePackage{booktabs}
\RequirePackage{colortbl}
\RequirePackage{tabularx}
\RequirePackage{makecell}
\RequirePackage{pifont}
\RequirePackage{arydshln}
\usepackage{tikz, tabularx}
\RequirePackage{amsmath,amsfonts}
\RequirePackage{algorithmic}
\RequirePackage{algorithm}
\RequirePackage{array}
\RequirePackage{subcaption}
\usepackage{enumitem}
\usepackage{glossaries}
\usepackage{tikz}
\usetikzlibrary{shapes.geometric, arrows.meta, positioning, fit, backgrounds, calc}
\usepackage{amsmath}  

\usepackage{colortbl}
\definecolor{rowgray}{gray}{0.93}

\usepackage{xcolor}

\RequirePackage{url}
\RequirePackage{verbatim}
\RequirePackage{graphicx}
\definecolor{chain0fill}{RGB}{219,234,245}
\definecolor{chain0draw}{RGB}{49,110,160}
\definecolor{chain1fill}{RGB}{220,232,220}
\definecolor{chain1draw}{RGB}{60,110,70}
\definecolor{occ_fill}{RGB}{224,232,242}
\definecolor{free_fill}{RGB}{245,240,230}
\definecolor{hdr_fill}{RGB}{238,238,238}
\definecolor{fifo_occ}{RGB}{253,245,215}
\definecolor{arr_head}{RGB}{180,20,20}
\definecolor{arr_fifo}{RGB}{100,50,140}
\definecolor{arr_c0}{RGB}{40,100,150}
\definecolor{arr_c1}{RGB}{50,100,55}

\RequirePackage{orcidlink}
\RequirePackage[noabbrev,capitalise]{cleveref}

\RequirePackage{cite}

\newcommand{\etal}{\emph{et al.}}
\newcommand{\x}{$\times$}

\renewcommand{\subsubsection}[1]{\paragraph*{\textbf{#1}}}

\DeclareSIUnit{\x}{\!\ensuremath{\times}}
\DeclareSIUnit\bit{b}
\DeclareSIUnit\GE{GE}
\DeclareSIUnit\kGE{\kilo\GE}
\DeclareSIUnit\MGE{\mega\GE}
\newlength\myheight
\newlength\mydepth
\settototalheight\myheight{Xygp}
\usepackage{eso-pic}
\usepackage{xcolor}

\AddToShipoutPictureBG*{%
  \AtPageUpperLeft{%
    \put(0,-28){%
      \makebox[\paperwidth][c]{%
        \footnotesize \color{darkgray} This article has been accepted for publication in \textit{IEEE Transactions on Very Large Scale Integration (VLSI) Systems}.
      }%
    }%
  }%
}

\begin{document}
\ifx\showrebuttal\undefined
    \newcommand{\rev}[1]{#1}
    \newcommand{\revdel}[1]{}
    \newcommand{\revrep}[2]{#2}
    \newcommand{\revprg}{}
\else
    \newcommand{\rev}[1]{\textcolor{blue}{#1}}
    \newcommand{\revdel}[1]{\textcolor{red}{\st{#1}}}
    \newcommand{\revrep}[2]{\revdel{#1} \rev{#2}}
    \newcommand{\revprg}{\hspace{-0.5ex}\textcolor{red}{\scalebox{.2}[1.5]{$\blacksquare$}}\hspace{-0.5ex}}
\fi
%
%


\title{Scalable AXI4 Transaction Monitoring for Mixed-Criticality SoCs: From Phase-Level Precision to ID-Level Efficiency}

\ifx\showrevision\undefined
    \newcommand{\todo}[1]{{#1}}
\else
    \newcommand{\todo}[1]{{\textcolor{red}{#1}}}
    \AddToShipoutPictureFG{%
        \put(%
            8mm,%
            \paperheight-1.5cm%
            ){\vtop{{\null}\makebox[0pt][c]{%
                \rotatebox[origin=c]{90}{%
                    \huge\textcolor{red!75}{\reviewpass}%
                }%
            }}%
        }%
    }
\fi

\author{
    Chaoqun~Liang\orcidlink{0009-0008-0556-7758},~\IEEEmembership{Graduate Student Member, IEEE},
    Thomas~Benz\orcidlink{0000-0002-0326-9676},~\IEEEmembership{Member, IEEE},
    Alessandro~Ottaviano\orcidlink{0009-0000-9924-3536},~\IEEEmembership{Member, IEEE},
    Michael Rogenmoser\orcidlink{0000-0003-4622-4862},~\IEEEmembership{Graduate Student Member, IEEE},
    Luca~Benini\orcidlink{0000-0001-8068-3806},~\IEEEmembership{Fellow, IEEE},
    Angelo~Garofalo\orcidlink{0000-0002-7495-6895},~\IEEEmembership{Member, IEEE},
    and~Davide~Rossi\orcidlink{0000-0002-0651-5393},~\IEEEmembership{Senior Member,~IEEE}
    \IEEEcompsocitemizethanks{%
    \IEEEcompsocthanksitem C.~Liang, L.~Benini, A.~Garofalo and D.~Rossi are with the Department of Electrical, Electronic, and Information Engineering (DEI), University of Bologna, Bologna, Italy.\protect\\
    E-mail: \{chaoqun.liang\}@unibo.it\protect\\
    T.~Benz, M.~Rogenmoser, L.~Benini and A.~Garofalo are with the Integrated Systems Laboratory (IIS), ETH Zurich, Switzerland.\protect\\
    T.~Benz is with lowRISC C.I.C., Cambridge, United Kingdom.\protect\\
    A.~Ottaviano is with Tenstorrent USA, Inc., Santa Clara, California.\protect\\
    }%
}

\markboth{}%
{Liang \MakeLowercase{\etal}: \title}

\maketitle

\glsresetall

\begin{abstract}
Mixed-criticality Systems-on-Chip (SoCs) with on-chip interconnects based on the AXI4 open standard protocol lack a protocol-level timeout mechanism, exposing systems to deadlocks and missed real-time deadlines when subordinate devices or managers fail or stall due to hardware faults, radiation-induced upsets, or software errors. \rev{This work presents a configurable hardware intellectual property (IP), non-intrusive in fault-free operation,} that detects AXI4 protocol violations and timing faults at runtime and restores interconnect liveness through a \textit{cut-and-drain} isolation mechanism. To address the fundamental trade-off between monitoring granularity and area cost, we introduce three designs at decreasing monitoring granularity: Phase-Level Tracking (PLT), which provides cycle-accurate fault localization across individual protocol phases; Channel-Level Tracking (CLT), which coalesces per-phase monitors into channel-level supervision; and ID-Level Tracking (ILT), which achieves sub-linear area scaling by monitoring only per-ID transaction boundaries. Synthesized in GlobalFoundries 12 nm technology, CLT reduces area by 36.7\% relative to PLT while preserving worst-case detection bounds at a minimal detection latency overhead, whereas ILT achieves an 89.2\% area reduction suitable for tightly constrained deployments at the cost of a $3.7\times$ higher median detection latency with coarser fault localization. Fault injection campaigns on a RISC-V SoC across 1.2 million scenarios confirm \rev{that no fault manifesting as an AXI4 protocol or liveness violation escaped detection}, with observed detection latencies consistently bounded by theoretical worst-case predictions.
\end{abstract}


\section{Introduction}   

Mixed-criticality System-on-Chip (SoC) architectures are increasingly adopted in automotive and aerospace embedded systems, where safety-critical functions such as vehicle control, flight management, and collision avoidance coexist on the same silicon platform as non-critical workloads such as diagnostic logging, data aggregation, and human-machine interface rendering ~\cite{b24,burns2017}. This integration demands not only high computational performance but also strict fault containment, temporal determinism, and compliance with functional safety standards such as ISO 26262~\cite{iso26262} in automotive~\cite{jiang2020mc} and DO-254~\cite{do254} in aerospace applications. Failures in these environments can have catastrophic consequences, hence standards mandate comprehensive fault detection, isolation, and recovery across all architectural layers. Fault resilience at the chip level has been studied through hardware diversity and redundancy approaches~\cite{resilogic}.

A fault in the system interconnect can propagate across the entire SoC, compromising both safety-critical and non-critical subsystems and potentially requiring a full system reset. The \rev{Advanced eXtensible Interface~4 (AXI4)}~\cite{b29}, defined in ARM's AMBA specification, has emerged as the dominant interconnect standard for high-performance safety-critical SoCs, yet it defines no mandatory timeout mechanism. When a subordinate device fails or stalls, its outstanding transactions might never complete. Since AXI4 imposes no deadline on subordinate response, the protocol cannot distinguish a slow subordinate from a permanently failed one. The resulting deadlock might propagate upstream,
potentially causing missed real-time deadlines, which is unacceptable in safety-critical operation. Critically, this failure mode cannot be prevented by formal verification or pre-silicon checking, as it arises from device-level failures at runtime~\cite{b14}. Real-time AXI scheduling~\cite{axiicrt} addresses prioritization but not runtime fault recovery.

Prior work has approached AXI4's lack of robustness when facing subordinate failure from several directions, each addressing a subset of the problem: protocol checkers~\cite{b6,amd_pc} enforce compliance rules but provide no timeout supervision or fault recovery; timeout mechanisms~\cite{b_xilinx_to,b11} can counter coarse-grained stalls but lack per-phase visibility and multi-transaction support; performance 
monitors~\cite{b13,b_ravi} target throughput characterization without fault detection capability; and access control solutions~\cite{b2,ims} address intentional security violations rather than unintentional faults. No existing solution simultaneously covers temporal 
supervision, fault containment, resource efficiency, and diagnostic observability under the area constraints of mixed-criticality SoCs.

Our previous work~\cite{b30} introduced the \rev{Transaction Monitoring Unit (TMU)}, a drop-in hardware IP that monitors AXI4 traffic between subordinate devices and the interconnect, detecting protocol violations, timing anomalies, and transaction stalls, and initiating recovery through AXI-compliant error response fabrication and hardware reset. However, the original TMU incurred significant area overhead ensuing from per-transaction counters and metadata replication, posing deployment challenges for resource-constrained SoCs with deep transaction concurrency.

An AXI4 \rev{TMU} must address four fundamental challenges: \textit{(C1)~Temporal Supervision}, detecting protocol and timing violations across interleaved, out-of-order transactions; \textit{(C2)~Fault Containment}, restoring interconnect liveness by injecting well-formed error responses before real-time deadlines are missed; \textit{(C3)~Resource Efficiency}, managing state for numerous outstanding transactions under strict area constraints; and \textit{(C4)~Diagnostic Observability}, providing sufficient observability to support post-fault root-cause analysis without prohibitive overhead.

This work addresses (C1)--(C4) through three TMU variants at decreasing monitoring granularity: Phase-Level Tracking (PLT), Channel-Level Tracking (CLT), and ID-Level Tracking (ILT), spanning the capability--complexity trade-off. Recovery is augmented with a \textit{cut-and-drain} isolation mechanism that immediately restores interconnect liveness upon fault detection.

The key contributions of this paper are:
\begin{itemize}
    \item An AXI4 TMU hardware IP, non-intrusive in fault-free operation, that detects protocol violations and timing faults from both managers and subordinates in bounded time, with guaranteed interconnect recovery upon fault detection.

    \item A three-tier monitoring architecture: PLT, CLT, and ILT, characterizing the trade-off between area cost, detection latency, and diagnostic capability, enabling designers to align monitoring overhead with subordinate criticality and safety certification requirements.

    \item Complete characterization of all three variants in terms of fault detection coverage and detection latency, and area scaling across various concurrency configurations in GlobalFoundries 12~nm technology.

   \item System-level fault injection campaigns on a RISC-V SoC, targeting both manager and subordinate IPs, confirming \rev{that no fault escaped detection within each variant's detectable class (Section~\ref{sec:faultmodel})}, with empirical detection latencies consistent with theoretical WCDT bounds.

\end{itemize}

\rev{The remainder of this paper is organized as follows.
Section~\ref{sec:overview} reviews the AXI4 ordering rules underlying the design and
defines the fault model. Section~\ref{sec:archi} presents the three TMU variants.
Section~\ref{sec:wcdt} derives the worst-case detection and liveness recovery bounds.
Section~\ref{sec:results} reports area, timing, power, and fault-injection
results. Section~\ref{sec:related} surveys related work, and
Section~\ref{sec:conclusion} concludes.}
\section{AXI4 Protocol Overview}
\label{sec:overview}
\subsection{Transactions, Phases, and Channels}
A \textit{transfer} is the atomic unit of AXI4 communication: a single beat of information exchanged over a \texttt{VALID}/\texttt{READY} handshake, completing on the rising clock edge when both signals are asserted. A \textit{transaction} is a sequence of transfers spanning the Write Address (AW), Write Data (W), and Write Response (B) channels for writes, or the Read Address (AR) and Read Data (R) channels for reads, with burst length given by \texttt{AWLEN}/\texttt{ARLEN}, the final beat marked by \texttt{WLAST}/\texttt{RLAST}, and completion status returned via \texttt{BRESP}/\texttt{RRESP}. 
In this work, a \textit{phase} denotes a distinct protocol event or transition within a transaction's lifecycle (address issuance, data beat transfer, or response reception), used as a semantic boundary for monitoring and timing measurement; the term is not part of the formal AXI4 specification~\cite{b29}.

A manager may issue multiple outstanding transactions, each tagged with an identifier (\texttt{AxID}) consistent across all its transfers, so transactions with different IDs
may complete out of order. Two asymmetries drive the TMU design: the W channel carries no transaction ID, so subordinates infer ownership from AW issue order, whereas each R beat explicitly identifies its owning transaction via \texttt{RID}.

\vspace{-6pt}
\subsection{Ordering Rules}
\label{sec:ordering_rules}
AXI4 defines the following ordering requirements, which directly determine
the TMU's tracking structures and counter allocation strategy
(Section~\ref{sec:archi}).

\textit{Write ordering:}
\begin{enumerate}[itemsep=0pt,topsep=2pt,parsep=0pt]
\item[\textit{W0.}] W-channel data must follow the same order as the corresponding AW transfers.
\item[\textit{W1.}] Transactions with different IDs may complete (B response) in any order.
\item[\textit{W2.}] Transactions sharing the same ID must be issued and completed in order.
\end{enumerate}

\textit{Read ordering:}
\begin{enumerate}[itemsep=0pt,topsep=2pt,parsep=0pt]
\item[\textit{R0.}] Read transactions with different IDs may return data in any order.
\item[\textit{R1.}] R-channel data beats may be interleaved across IDs; \texttt{RID} identifies the owning transaction.
\item[\textit{R2.}] Read transactions sharing the same ID must return data in issue order.
\end{enumerate}

\vspace{-6pt}
\subsection{Fault Model and Scope}
\label{sec:faultmodel}
The TMU observes AXI4 traffic at the interface between the interconnect and the monitored subordinate, so its detection scope is limited by what a fault manifests as at that interface. Faults perturbing protocol signalling or transaction progress are \emph{detectable}: stalled or
absent handshakes, unexpected or duplicate \texttt{BID}/\texttt{RID}, premature, missing or excess \texttt{WLAST}/\texttt{RLAST}, and burst-length mismatches. Faults leaving signalling well-formed are \emph{undetectable} by construction, principally silent data corruption,
which requires complementary end-to-end protection such as ECC. All coverage claims in this work refer to the detectable class, whose per-variant scope is given in Table~\ref{tab:tmu_comparison}.
\section{Architecture}
\label{sec:archi}

\begin{figure}[htbp]
    \centering
    \includegraphics[width=0.5\textwidth, height=0.23\textwidth]{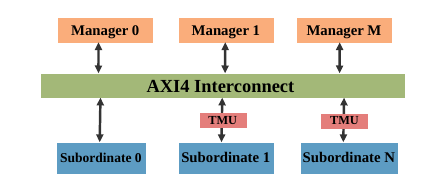}
    \vspace{-15pt}
    \caption{TMU placement in an AXI4-based SoC.}
    \label{fig:placement}
\end{figure}

\begin{table*}[t]
\centering
\caption{Comparison of PLT, CLT, and ILT across monitoring granularity, area complexity, and diagnostic capability.}
\resizebox{\textwidth}{!}{%
\begin{tabular}{|l|c|c|c|}
\hline
\textbf{TMU Variant} &
  \textbf{Phase-Level Tracking (PLT)} &
  \textbf{Channel-Level Tracking (CLT)} &
  \textbf{ID-Level Tracking (ILT)} \\ \hline
\textbf{Detection Granularity} &
  Per-phase within transaction &
  Per-channel liveness &
  Per-ID transaction boundary \\ \hline
\textbf{Timeout Counters} &
  \begin{tabular}[c]{@{}c@{}}
    6 write (C\textsubscript{w0}--C\textsubscript{w5}) + 4 read (C\textsubscript{r0}--C\textsubscript{r3a})\\
    {\footnotesize Separate address, data, and response phase thresholds}\\
    {\footnotesize + beat count validation (C\textsubscript{w3b}/C\textsubscript{r3b})}
  \end{tabular} &
  \begin{tabular}[c]{@{}c@{}}
    3 write (C\textsubscript{w0}--C\textsubscript{w2}) + 2 read (C\textsubscript{r0}--C\textsubscript{r1})\\
    {\footnotesize One threshold per channel (AW/W/B or AR/R)}\\
    {\footnotesize + beat count validation}
  \end{tabular} &
  \begin{tabular}[c]{@{}c@{}}
    2 write (C\textsubscript{w0}--C\textsubscript{w1}) + 2 read (C\textsubscript{r0}--C\textsubscript{r1})\\
    {\footnotesize One threshold per ID;}\\
    {\footnotesize end-to-end inter-completion boundaries}
  \end{tabular} \\ \hline
\textbf{Area Complexity} &
  $\mathcal{O}(N_{ID} \times D)$ &
  $\mathcal{O}(N_{ID} \times D)$ &
  $\mathcal{O}(N_{ID})$ \\ \hline
\textbf{Error Detection Response} &
  Immediate (per-phase threshold) &
  Immediate (per-channel threshold) &
  Slow (end-to-end threshold) \\ \hline
\hline
\multicolumn{4}{|c|}{\textbf{Diagnostic Capability}} \\ \hline
Faulting transaction identifier           & $\checkmark$ & $\checkmark$ & $\checkmark$ \\ \hline
Fault direction (write / read)            & $\checkmark$ & $\checkmark$ & $\checkmark$ \\ \hline
Fault type (timeout / protocol violation) & $\checkmark$ & $\checkmark$ & $\checkmark$ \\ \hline
Fault source (manager / subordinate)      & $\checkmark$ & \rev{$\circ$~{\footnotesize(address phase)}} & \rev{$\circ$~{\footnotesize(address phase)}} \\ \hline
Fault detection latency (cycles)          & $\checkmark$ & $\checkmark$ & $\checkmark$ \\ \hline
Burst beats transferred                   & $\checkmark$ & $\checkmark$ & $\times$   \\ \hline
Faulting counter resolution               & $\checkmark$~{\footnotesize(C\textsubscript{w0}--C\textsubscript{w5}/C\textsubscript{r0}--C\textsubscript{r3b})} & $\circ$~{\footnotesize(per channel)} & $\times$
\\ \hline
\hline
\multicolumn{4}{|c|}{\textbf{Protocol Violation Detection}} \\ \hline
Unexpected / duplicate BID or RID         & $\checkmark$ & $\checkmark$ & $\checkmark$ \\ \hline
Burst-length mismatch                     & $\checkmark$ & $\checkmark$ & $\times$   \\ \hline
\end{tabular}%
}
\begin{tablenotes}
\footnotesize
\item $N_{ID}$: number of distinct outstanding AXI transaction IDs supported by the TMU.\quad
      $D$: maximum outstanding transactions per ID (outstanding depth).\\
      $\checkmark$~= supported;\quad
      \rev{$\circ$~= partially supported;}\quad
      $\times$~= not supported.
\end{tablenotes}
\vspace{-10pt}
\label{tab:tmu_comparison}
\end{table*}

The \rev{TMU} is a dedicated hardware IP that \rev{observes in-flight AXI4 transactions without adding arbitration, buffering, or protocol state to the AXI4 path and requires no modification to the crossbar microarchitecture.} It provides real-time monitoring and time-predictable fault detection for managers and subordinates in mixed-criticality AXI4 SoCs. As shown
in Figure~\ref{fig:placement}, one TMU instance is placed per monitored subordinate, ensuring all managers targeting that subordinate are observed without per-manager hardware replication. The TMU examines both request and
response channels to detect bidirectional protocol and timing violations, whether originating in the subordinate or induced by managers.

The TMU delivers four core functions: (1)~\textit{Protocol Violation Detection} scans all five AXI4 channels to identify incomplete handshakes, unmatched response identifiers, and burst-length mismatches; if left undetected, such violations can cause unrecoverable deadlocks requiring global reset; (2)~\textit{Time-bounded Fault Detection} enforces user-configurable timeout thresholds on transaction progress through a set of dedicated counters, each incrementing each cycle in the absence of the expected protocol event and flagging a fault upon threshold expiry. Monitoring granularity ranges from per-phase supervision in PLT to end-to-end transaction boundaries in ILT (Table~\ref{tab:tmu_comparison}); (3)~\textit{Fault Management and Recovery} distinguishes subordinate-induced faults from manager-induced ones and asserts a hardware interrupt to notify the processor. For subordinate-induced faults, the TMU additionally asserts a reset request to an external hardware reset controller \rev{connected} to the subordinate device. Concurrently, the TMU executes \textit{cut-and-drain} isolation: the \textit{cut} phase blocks all subsequent requests to the faulting subordinate, while the \textit{drain} phase retires all pending transactions with synthesized error responses to prevent interconnect deadlock (Section~\ref{sec:plt:fault}); and (4)~\textit{Statistics and Debug Support} captures and logs diagnostic metadata for post-fault root-cause analysis (Section~\ref{sec:plt:fault}).

\begin{figure}[!t]
\centerline{\includegraphics[width=0.52\textwidth, height=0.35\textwidth]{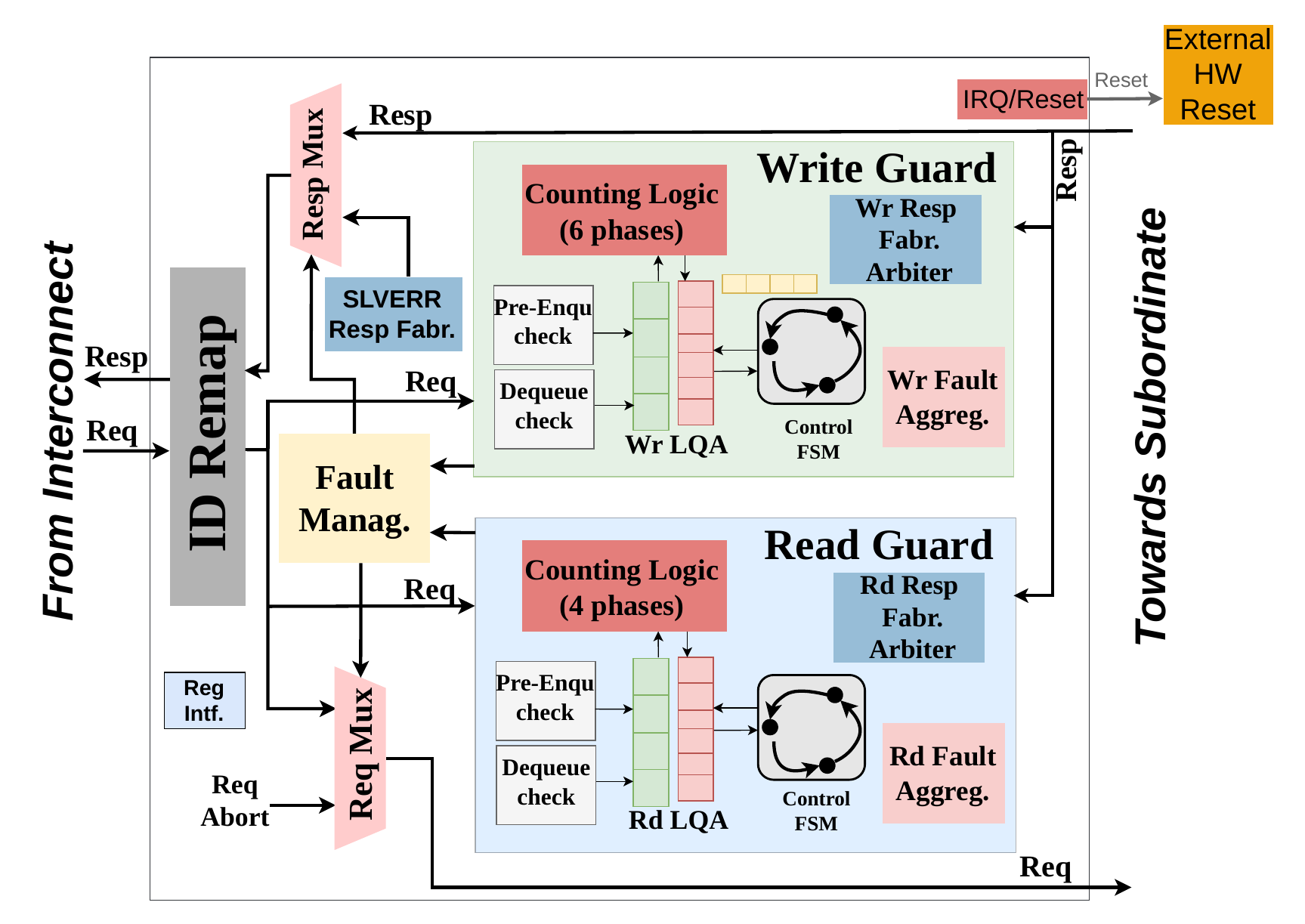}}
\caption{TMU Phase-Level Tracking (PLT) architecture showing independent Write Guard and Read Guard paths.}
\label{fig:plt_top}
\end{figure}

Implementing these functions entails a fundamental trade-off between monitoring granularity and area overhead. Fine-grained monitoring requires per-transaction state, scaling with both the number of supported IDs and the maximum outstanding depth per ID. High-end SoCs may tolerate the TMU area overhead implied by full per-transaction tracking for deep observability, but tightly resource-constrained SoCs must monitor multiple subordinates under strict area and power budgets~\cite{b26}. Moreover, criticality and role vary across subordinates, making a uniform monitoring policy suboptimal. We therefore design the TMU as a family of three variants: \textit{Phase-Level Tracking (PLT)}, \textit{Channel-Level Tracking (CLT)}, and \textit{ID-Level Tracking (ILT)}, spanning the capability-complexity trade-off (Table~\ref{tab:tmu_comparison}). Crucially, these variants are not mutually exclusive: a single SoC design may deploy different TMU variants simultaneously across subordinates, assigning PLT to safety-critical components requiring precise fault localization, CLT to standard components balancing cost and observability, and ILT to area-constrained endpoints where lightweight liveness monitoring suffices.

\begin{figure}[htbp]
\centering
\definecolor{lqa_c0}{RGB}{49,110,160}
\definecolor{lqa_c1}{RGB}{60,110,70}
\definecolor{lqa_c2}{RGB}{170,120,50}
\definecolor{lqa_occ_fill}{RGB}{224,232,242}
\definecolor{lqa_free_fill}{RGB}{245,240,230}
\definecolor{lqa_hdr_fill}{RGB}{238,238,238}
\definecolor{lqa_fifo_occ}{RGB}{253,245,215}
\definecolor{lqa_arr_head}{RGB}{180,20,20}
\definecolor{lqa_arr_fifo}{RGB}{100,50,140}

\begin{tikzpicture}[
    scale=0.56, transform shape,
    bucket/.style={
        rectangle, rounded corners=5pt, draw=black!70, line width=0.8pt,
        minimum width=3.4cm, minimum height=2.35cm, align=left, font=\normalsize,
        fill=lqa_occ_fill},
    bucket_free/.style={bucket, fill=lqa_free_fill, draw=black!50},
    txn_base/.style={
    rectangle, rounded corners=5pt, line width=0.8pt,
    minimum width=2.8cm, minimum height=1.9cm, align=left,
    inner xsep=6pt, font=\small},
    txn_free/.style={txn_base, fill=lqa_free_fill, draw=black!50},
    txn_id0/.style={txn_base, fill=lqa_occ_fill, draw=lqa_c0, line width=1.6pt},
    txn_id1/.style={txn_base, fill=lqa_occ_fill, draw=lqa_c1, line width=1.6pt},
    txn_id2/.style={txn_base, fill=lqa_occ_fill, draw=lqa_c2, line width=1.6pt},
    fifo_cell/.style={
        rectangle, rounded corners=3pt, draw=black!70, line width=0.8pt,
        minimum width=2.2cm, minimum height=0.85cm, font=\normalsize,
        fill=lqa_fifo_occ},
    fifo_empty/.style={fifo_cell, fill=white, draw=black!40},
    label_box/.style={
        rectangle, draw=black!80, line width=1pt, fill=lqa_hdr_fill,
        font=\normalsize\bfseries, minimum height=0.65cm, rounded corners=3pt},
    annotation/.style={font=\small, text=black!70},
    head_arrow/.style={
        -{Stealth[length=3mm, width=2mm]}, color=lqa_arr_head, line width=1.2pt},
    tail_arrow/.style={
        -{Stealth[length=3mm, width=2mm]}, color=lqa_arr_head!70, line width=1pt, dashed},
    next_c0/.style={
        -{Stealth[length=2.5mm, width=1.8mm]}, color=lqa_c0, line width=1.2pt},
    next_c1/.style={
        -{Stealth[length=2.5mm, width=1.8mm]}, color=lqa_c1, line width=1.2pt},
    fifo_arrow/.style={
        -{Stealth[length=2.5mm, width=1.8mm]}, color=lqa_arr_fifo, line width=1.1pt},
]

\node[label_box] (idq_label) at (0,0) {ID Queue (Head-Tail Buckets)};

\node[bucket, below=0.4cm of idq_label] (bucket0) {
    \textbf{Bucket 0}\\[1pt]
    \texttt{ID}: 0\quad \texttt{free}: \textcolor{red!80!black}{\textbf{$\times$}}\\
    \texttt{head}: slot\,0\\
    \texttt{tail}: slot\,4\\
    \texttt{C\textsubscript{w4}}, \texttt{C\textsubscript{w5}}: \textit{active}
};
\node[bucket_free, below=0.25cm of bucket0] (bucket1) {
    \textbf{Bucket 1}\\[1pt]
    \texttt{ID}: {---}\quad \texttt{free}: \textcolor{green!50!black}{\textbf{$\surd$}}\\
    \texttt{head}: {---}\\
    \texttt{tail}: {---}\\
    \texttt{C\textsubscript{w4}}, \texttt{C\textsubscript{w5}}: \textit{---}
};
\node[bucket, below=0.25cm of bucket1] (bucket2) {
    \textbf{Bucket 2}\\[1pt]
    \texttt{ID}: 2\quad \texttt{free}: \textcolor{red!80!black}{\textbf{$\times$}}\\
    \texttt{head}: slot\,2\\
    \texttt{tail}: slot\,2\\
    \texttt{C\textsubscript{w4}}, \texttt{C\textsubscript{w5}}: \textit{idle}
};
\node[bucket, below=0.25cm of bucket2] (bucket3) {
    \textbf{Bucket 3}\\[1pt]
    \texttt{ID}: 1\quad \texttt{free}: \textcolor{red!80!black}{\textbf{$\times$}}\\
    \texttt{head}: slot\,3\\
    \texttt{tail}: slot\,5\\
    \texttt{C\textsubscript{w4}}, \texttt{C\textsubscript{w5}}: \textit{active}
};
\node[annotation, below=0.1cm of bucket3, text width=3.4cm, align=center] {
    Per-ID metadata (\textit{MaxUniqIDs} entries)
};

\node[label_box] (txn_label) at (5.2,0) {Transaction Table};

\node[txn_id0, below=0.4cm of txn_label] (slot0) {
    \textbf{slot\,0}\\[1pt]
    \texttt{ht\_idx}: 0\quad \texttt{free}: \textcolor{red!80!black}{\textbf{$\times$}}\\
    \texttt{next}: slot\,4\\
    \texttt{beats}: 32\\
    \texttt{state}: W\_RESP
};
\node[txn_free, below=0.18cm of slot0] (slot1) {
    \textbf{slot\,1}\\[1pt]
    \texttt{ht\_idx}: {---}\quad \texttt{free}: \textcolor{green!50!black}{\textbf{$\surd$}}\\
    \texttt{next}: {---}\\
    \texttt{beats}: {---}\\
    \texttt{state}: {---}
};
\node[txn_id2, below=0.18cm of slot1] (slot2) {
    \textbf{slot\,2}\\[1pt]
    \texttt{ht\_idx}: 2\quad \texttt{free}: \textcolor{red!80!black}{\textbf{$\times$}}\\
    \texttt{next}: {---}\\
    \texttt{beats}: 16\\
    \texttt{state}: W\_DATA
};
\node[txn_id1, below=0.18cm of slot2] (slot3) {
    \textbf{slot\,3}\\[1pt]
    \texttt{ht\_idx}: 3\quad \texttt{free}: \textcolor{red!80!black}{\textbf{$\times$}}\\
    \texttt{next}: slot\,5\\
    \texttt{beats}: 8\\
    \texttt{state}: W\_RESP
};
\node[txn_id0, below=0.18cm of slot3] (slot4) {
    \textbf{slot\,4}\\[1pt]
    \texttt{ht\_idx}: 0\quad \texttt{free}: \textcolor{red!80!black}{\textbf{$\times$}}\\
    \texttt{next}: {---}\\
    \texttt{beats}: 64\\
    \texttt{state}: W\_ADDR        
};
\node[txn_id1, below=0.18cm of slot4] (slot5) {
    \textbf{slot\,5}\\[1pt]
    \texttt{ht\_idx}: 3\quad \texttt{free}: \textcolor{red!80!black}{\textbf{$\times$}}\\
    \texttt{next}: {---}\\
    \texttt{beats}: 4\\
    \texttt{state}: W\_ADDR
};
\node[annotation, below=0.1cm of slot5, text width=4.5cm, align=center] {
    Flat storage pool (\textit{MaxOutstdTxns} entries)
};

\node[label_box] (fifo_label) at (10.8,0) {W-Ownership FIFO};

\node[fifo_cell, below=0.4cm of fifo_label, align=center,
      draw=lqa_arr_head, line width=1.2pt] (fifo0) {
    \textbf{HEAD} $\to$ slot\,2\\[-1pt]
    {\small\textcolor{lqa_arr_head}{(current W owner)}}
};
\node[fifo_cell, below=0.12cm of fifo0] (fifo1) {slot\,4};
\node[fifo_cell, below=0.12cm of fifo1] (fifo2) {slot\,5};
\node[fifo_empty, below=0.12cm of fifo2] (fifo3) {\textit{(empty)}};
\node[fifo_empty, below=0.12cm of fifo3, draw=black!40] (fifo4) {\textbf{TAIL}};

\node[annotation, below=0.1cm of fifo4, text width=3.3cm, align=center] {
    W-channel serialization\\(\textit{MaxOutstdTxns} depth)
};

\draw[head_arrow] ([yshift=3pt]bucket0.east) -- ++(0.20,0) |- (slot0.west);
\draw[tail_arrow] ([yshift=-5pt]bucket0.east) -- ++(0.38,0) |- (slot4.west);

\draw[head_arrow] (bucket2.east) -- ++(0.20,0) |- (slot2.west);

\draw[head_arrow] ([yshift=3pt]bucket3.east) -- ++(0.30,0) |- (slot3.west);
\draw[tail_arrow] ([yshift=-5pt]bucket3.east) -- ++(0.48,0) |- (slot5.west);

\draw[next_c0] ([yshift=5pt]slot0.east) -- ++(0.45,0) |- ([yshift=5pt]slot4.east);
\node[font=\small\bfseries, text=lqa_c0, fill=white, inner sep=1pt] at ([xshift=0.45cm]$(slot0.east)!0.45!(slot4.east)$) {\textit{next}};
\draw[next_c1] ([yshift=5pt]slot3.east) -- ++(0.65,0) |- ([yshift=5pt]slot5.east);
\node[font=\small\bfseries, text=lqa_c1, fill=white, inner sep=1pt] at ([xshift=0.65cm]$(slot3.east)!0.55!(slot5.east)$) {\textit{next}};
\draw[fifo_arrow] (fifo0.west) -- ++(-0.45,0) |- ([yshift=-5pt]slot2.east);
\draw[fifo_arrow] (fifo1.west) -- ++(-0.65,0) |- ([yshift=-5pt]slot4.east); 
\draw[fifo_arrow] (fifo2.west) -- ++(-0.85,0) |- ([yshift=-5pt]slot5.east);  
\path (idq_label.west) -- (fifo_label.east) coordinate[midway] (fig_center);
\node[draw=black!60, line width=0.6pt, rounded corners=3pt,
      inner xsep=0.12cm, inner ysep=0.1cm,
      fill=white, font=\small] (legend_box) at ([yshift=-1.4cm]fig_center |- slot5.south) {
    \begin{tabular}{@{}l@{\hskip 0.35cm}l@{\hskip 0.35cm}l@{\hskip 0.35cm}l@{\hskip 0.35cm}l@{}}
        \tikz\draw[lqa_arr_head, line width=1pt, -{Stealth[length=1.6mm]}] (0,0) -- (0.35,0);~Head &
        \tikz\draw[lqa_arr_head!70, line width=0.8pt, dashed, -{Stealth[length=1.6mm]}] (0,0) -- (0.35,0);~Tail &
        \tikz\draw[lqa_arr_fifo, line width=0.9pt, -{Stealth[length=1.6mm]}] (0,0) -- (0.35,0);~FIFO ref &
        \tikz\node[fill=lqa_free_fill, draw=black!50, line width=0.4pt, minimum width=0.22cm, minimum height=0.16cm, rounded corners=1pt] {};~Free \\[0.08cm]
        \tikz\node[fill=lqa_occ_fill, draw=lqa_c0, line width=1pt, minimum width=0.22cm, minimum height=0.16cm, rounded corners=1pt] {};~ID\,=\,0 &
        \tikz\node[fill=lqa_occ_fill, draw=lqa_c1, line width=1pt, minimum width=0.22cm, minimum height=0.16cm, rounded corners=1pt] {};~ID\,=\,1 &
        \tikz\node[fill=lqa_occ_fill, draw=lqa_c2, line width=1pt, minimum width=0.22cm, minimum height=0.16cm, rounded corners=1pt] {};~ID\,=\,2 &
        \multicolumn{2}{@{}l@{}}{\texttt{C\textsubscript{w4}},\texttt{C\textsubscript{w5}}: per-ID counters} \\
    \end{tabular}
};

\end{tikzpicture}
\caption{Write Guard LQA structure.}
\label{fig:lqa_structure}
\vspace{-6pt}
\end{figure}
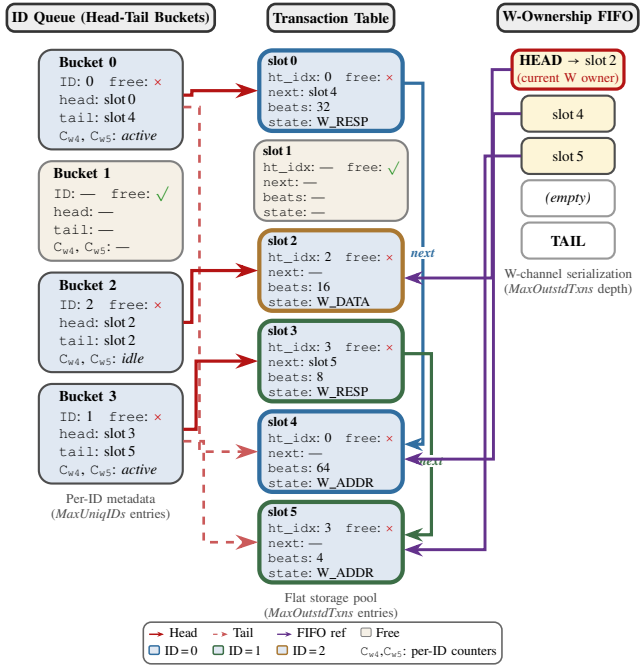

\rev{The three variants are presented incrementally. PLT is described in full, establishing the tracking structures and counter allocation from which the other two are derived; CLT and ILT are then described by what they retain, coalesce,
or eliminate relative to PLT, with reference back to the corresponding PLT components throughout.}

\vspace{-5pt}
\subsection{Phase-Level Tracking (PLT)}
\label{sec:plt}
PLT offers the finest monitoring granularity by decomposing each AXI4 transaction into discrete protocol phases, each with independent timeout supervision. This decomposition enables precise fault localization to the address handshake, data transfer, or response completion and supports comprehensive diagnostic logging. Achieving phase-level monitoring under AXI4's concurrent, out-of-order transaction model requires dedicated structures for ID namespace management, transaction tracking with ordering enforcement, and write-channel serialization. Figure~\ref{fig:plt_top} illustrates the resulting architecture: an \textit{ID Remapper} for ID namespace compression, independent monitoring paths where the \textit{Write Guard} supervises six write transaction phases and the \textit{Read Guard} supervises four read transaction phases, a shared \textit{Fault Management} block for fault isolation and recovery, and a \textit{Register Interface} for user-configuration and fault logging as described below. The following subsections trace a transaction's lifetime from ingress to completion or fault isolation.

\subsubsection{ID Remapper}
\label{sec:plt:id_remapper}
At TMU ingress, the ID Remapper compresses the external AXI4 ID. AXI4 ID widths are implementation-defined (typically 4--16~bits) and may be widened by the interconnect for routing, yet only a sparse subset of the namespace is active at any time. The remapper maps active IDs to a dense internal space bounded by \texttt{MaxUniqIDs}, reducing storage and operand width throughout the TMU. Only ID fields are compressed at ingress and decompressed at egress; all other signals pass through transparently. When active IDs reach capacity, new transactions are stalled by deasserting \texttt{AWREADY}/\texttt{ARREADY} until an ID is released. In practice, \texttt{MaxUniqIDs} is configured at design time to match the number of unique IDs the connected manager can issue, so capacity exhaustion does not occur under normal operation.

\begin{table}[t]
  \centering
  \renewcommand{\arraystretch}{1.15}
  \caption{PLT phase counters: allocation strategy and diagnostic implication.}
  \label{tab:plt_cnters}
  \small
  \begin{tabular}{@{}llp{6.2cm}@{}}
    \toprule
    \textbf{Cnt.} & \textbf{Alloc.} & \textbf{Diagnostic Implication} \\
    \midrule
    \multicolumn{3}{@{}l}{\textit{Write path}} \\
    \midrule
    C\textsubscript{w0} & Shared\textsuperscript{a} & AW channel subordinate unresponsiveness \\
    C\textsubscript{w1} & Shared\textsuperscript{b} & Manager data path delay post-AW handshake \\
    C\textsubscript{w2} & Shared\textsuperscript{b} & Subordinate contention on first W beat \\
    C\textsubscript{w3a} & Shared\textsuperscript{b} & W channel stall between consecutive beats \\
    C\textsubscript{w3b} & Shared\textsuperscript{b} & Premature, missing, or excess WLAST \\
    C\textsubscript{w4} & Per-ID\textsuperscript{c} & Subordinate B response delay; unexpected or missing BID under out-of-order completion \\
    C\textsubscript{w5} & Per-ID\textsuperscript{c} & Manager backpressure on B handshake \\
    \midrule
    \multicolumn{3}{@{}l}{\textit{Read path}} \\
    \midrule
    C\textsubscript{r0} & Shared\textsuperscript{a} & AR channel subordinate unresponsiveness \\
    C\textsubscript{r1} & Per-ID\textsuperscript{c} & Subordinate data path delay post-AR handshake \\
    C\textsubscript{r2} & Per-ID\textsuperscript{c} & Manager contention on first R beat \\
    C\textsubscript{r3a} & Per-ID\textsuperscript{c} & R channel stall between consecutive beats; interleave-safe \\
    C\textsubscript{r3b} & Per-ID\textsuperscript{c} & Premature, missing, or excess RLAST \\
    \bottomrule
    \multicolumn{3}{@{}l}{\textsuperscript{a}Pre-enqueue, shared across all IDs} \\
    \multicolumn{3}{@{}l}{\textsuperscript{b}Shared, W-FIFO head only} \\
    \multicolumn{3}{@{}l}{\textsuperscript{c}Per-ID, ID Queue bucket} \\
  \end{tabular}
  \vspace{-6pt}
\end{table}

\subsubsection{Pre-Enqueue Monitoring}
\label{sec:plt:preenqueue}

To prevent resource exhaustion, the TMU employs a deferred allocation strategy. Since \texttt{AW}/\texttt{AR} handshakes may stall arbitrarily before the subordinate asserts \texttt{READY}, pre-allocating tracking slots at \texttt{AWVALID}/\texttt{ARVALID} assertion would exhaust tracking capacity under sustained backpressure from slow or faulty subordinates. Pre-enqueue monitoring therefore uses a dedicated lightweight watchdog, $C_{w0}$ for writes and $C_{r0}$ for reads (Table~\ref{tab:plt_cnters}), that monitors address handshake latency without consuming any tracking slot. If the subordinate fails to assert \texttt{READY} within the configured threshold, the TMU immediately raises a fault. The transaction tracking infrastructure presented next is engaged only after \texttt{AW}/\texttt{AR} handshake completion, ensuring that requests which may never be accepted cannot prematurely exhaust available slots.

\subsubsection{Linked Queue Architecture}
\label{sec:plt:lqa}


Upon successful pre-enqueue check, the transaction enters the Linked Queue Architecture (LQA). AXI4's concurrency and ordering rules (Section~\ref{sec:ordering_rules}) impose conflicting constraints: within each ID, responses complete in issue order (Rules~W2/R2), yet across IDs, completions arrive in arbitrary sequence (Rules~W1/R0). Similarly, W-channel data follows a single global serialization order (Rule~W0), while R-channel beats interleave freely across IDs (Rule~R1). The LQA resolves these constraints through per-ID linked chains, W-channel serialization by AW-issue order, and interleave-tolerant read tracking.

A static allocation reserving fixed entries per ID wastes area under asymmetric traffic, where some IDs have many outstanding transactions while others have few or none. PLT therefore introduces the LQA as depicted in Figure~\ref{fig:lqa_structure}, 
which dynamically allocates from a shared pool of transaction slots, decoupling total capacity (\texttt{MaxOutstdTxns}) from per-ID depth (\texttt{TxnPerUniqID}) so any ID can utilize available slots on demand. The LQA tracks state at two granularities: per-transaction (phase, beat progress, ID-chain linkage) and per-ID (ordering enforcement and timeout counters). It comprises three components: a \textit{Transaction Table}, an 
\textit{ID Queue}, and a \textit{W-Ownership FIFO}. The Read Guard uses identical Table and Queue structures but omits the W-FIFO, since R-channel beats are disambiguated by \texttt{RID} tags.

\textit{ID Queue (Head-Tail Buckets)} holds per-ID metadata in \texttt{MaxUniqIDs} entries, each storing a remapped ID, \texttt{head}/\texttt{tail} pointers to the oldest and newest Transaction Table slots in the chain, a \texttt{free} flag, and phase-specific timeout counters. Each ID maintains independent \texttt{head}/\texttt{tail} pointers, so out-of-order responses across different IDs (Rules~W1/R0) are handled naturally without cross-ID interference. In the Write Guard, each bucket stores response-phase counters $C_{w4}$/$C_{w5}$ to handle out-of-order B-completion behavior (Rule~W1). In the Read Guard, buckets store data-phase counters $C_{r1}$--$C_{r3b}$ per ID, enabling R-beat monitoring under arbitrary interleaving (Rule~R1).

\textit{Transaction Table} is a flat pool of \texttt{MaxOutstdTxns} entries, where any slot may hold any outstanding transaction regardless of ID. Each slot maintains five fields: the index of its owning ID Queue bucket (\texttt{ht\_idx}), a same-ID chain pointer (\texttt{next}), the expected beat count (\texttt{beats} = \texttt{AWLEN}+1 or \texttt{ARLEN}+1), a \texttt{free} flag, and the current protocol phase (\texttt{state}), which is updated by the TMU at each protocol transition as the transaction progresses through its lifecycle. Slots are allocated via parallel free-entry search on \texttt{AW}/\texttt{AR} handshake and released on response completion or fault. As Figure~\ref{fig:lqa_structure} illustrates, per-ID chains are formed via \texttt{next} pointers across physically non-adjacent slots, decoupling logical ordering from physical storage position.

\textit{W-Ownership FIFO} tracks which transaction currently holds ownership of the W-data channel. Rule~W0 requires W-data to follow \texttt{AW}-handshake order, yet the W-channel carries no ID tag; the subordinate infers ownership solely from \texttt{AW} issue order. The FIFO encodes this ordering: on \texttt{AW} handshake the new Transaction Table slot index is enqueued into the W-Ownership FIFO, and the FIFO head always identifies the current W-data owner. Subsequent transactions wait until \texttt{WLAST} of the preceding transaction is observed. No analogous structure is needed in the Read Guard, where \texttt{RID} tags allow direct routing of each R beat to the correct ID Queue bucket without serialization.

TMU capacity is parameterized at design time (Table~\ref{tab:capacity-table}): \texttt{MaxUniqIDs} sets the number of ID Queue buckets, \texttt{TxnPerUniqID} the depth of each bucket, and \texttt{MaxOutstdTxns} the W-Ownership FIFO depth bounding total outstanding write transactions.

\begin{table}[t]
\centering
\small
\caption{TMU structural capacity parameters.}
\label{tab:capacity-table}
\begin{tabular}{lp{4.6cm}l}
\toprule
\textbf{Parameter} & \textbf{Description} \\
\midrule
\textit{MaxUniqIDs}    & Number of ID Queue buckets; one per tracked ID \\
\addlinespace
\textit{TxnPerUniqID}  & Chain depth per ID bucket; bounds per-ID outstanding transactions \\
\addlinespace
\textit{MaxOutstdTxns} & W-Ownership FIFO depth; sets total outstanding write transactions \\
\bottomrule
\end{tabular}
\vspace{-4pt}
\end{table}

\textit{Enqueue Operation.} On \texttt{AW}/\texttt{AR} handshake, the TMU allocates a free Transaction Table slot via parallel search and populates it with the transaction's metadata. Simultaneously, an associative lookup matches \texttt{AWID}/\texttt{ARID} against occupied ID Queue buckets. If a matching bucket is found, the new slot is appended to the existing chain via the tail's \texttt{next} pointer and the tail advances, thus preserving issue order within the ID. If no bucket exists, a new ID Queue entry is allocated with \texttt{head} and \texttt{tail} both pointing to the new slot. For writes, the slot index is also enqueued into the W-Ownership FIFO. Figure~\ref{fig:lqa_structure} illustrates the resulting state: ID~0 forms the chain slot~0$\rightarrow$slot~4, ID~1 forms slot~3$\rightarrow$slot~5, and ID~2 holds a single-slot chain at slot~2. \rev{Continuing from this state, a write transaction with \texttt{AWID}$\,=1$
proceeds as follows. On the AW handshake the free-entry search returns slot~1,
the only free entry; the associative lookup matches bucket~3, which already
holds ID~1; slot~1 is appended to that chain through the \texttt{next} field of
the current tail, slot~5; and slot~1 is enqueued into the W-Ownership FIFO
behind slot~5. The transaction must now await the \texttt{WLAST} of the three
transactions ahead of it in the FIFO before transferring W data, per Rule~W0,
while its B response may complete out of order with respect to buckets~0 and~2,
per Rule~W1. A read follows the same path in the Read Guard but omits the FIFO
step, since \texttt{RID} identifies each R~beat's owner directly; its data-phase counters therefore reside in its own bucket and are immune to interleaving from other IDs, per Rule~R1.}

\subsubsection{Phase Counter Organization}
\label{sec:plt:counters}

\begin{figure}[htbp]
\centerline{\includegraphics[width=0.52\textwidth]{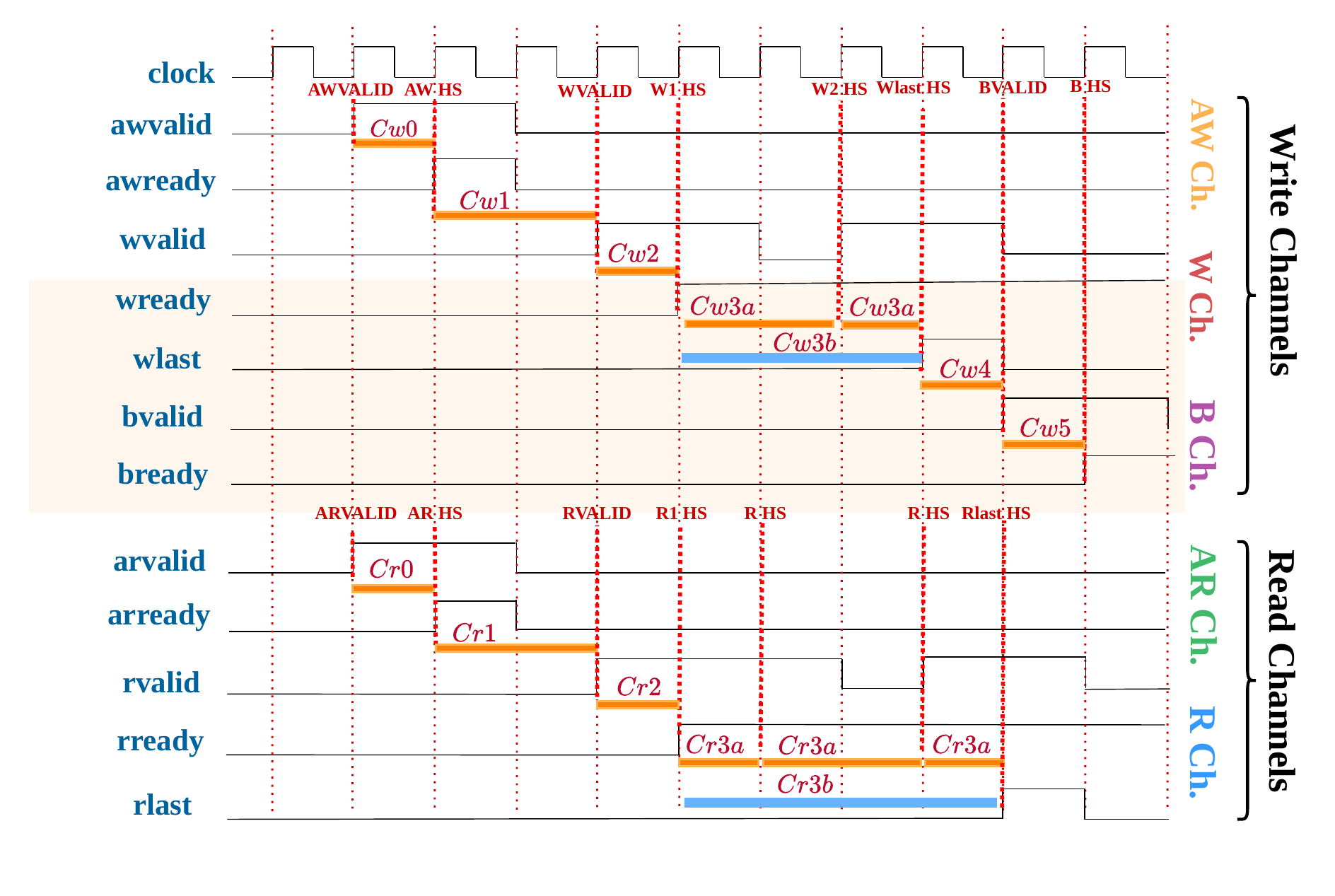}}
\vspace{-12pt}
\caption{Phase-level counters in PLT.}
\label{fig:plt_phases}
\vspace{-6pt}
\end{figure}

Once enqueued, each transaction is supervised by dedicated phase counters. Figure~\ref{fig:plt_phases} shows monitored intervals across write and read transactions, with counter spans aligned to the corresponding signal transitions. 
Table~\ref{tab:plt_cnters} maps each counter to its allocation and diagnostic implication.

\textit{Counter Allocation Strategy.} PLT exploits the asymmetry between write and read ordering rules to minimize area overhead. For the W-channel, Rule~W0 serialization means only one transaction holds W-data ownership at any time; PLT therefore maintains one shared set of data-phase counters $C_{w1}$ to $C_{w3b}$ assigned to the FIFO head (Shared\textsuperscript{b}). When a transaction is promoted to head, counters reset and bind to its progress; after \texttt{WLAST} they reassign to the next FIFO entry. This time-multiplexing enables full per-phase supervision without per-transaction counter replication.

For the R-channel, Rule~R1 interleaving across IDs would cause a single shared counter to be prematurely reset by beats from other IDs; each active ID therefore maintains its own independent data-phase counters (Per-ID\textsuperscript{c}), instantiated in the respective ID Queue bucket. Within a given ID, Rule~R2 prevents same-ID beat interleaving, so only one transaction per ID is in the data phase at a time; the same per-ID counter set is thus reused temporally for successive transactions without replication.

Within the data phase, the counter pair $C_{w3a}$/$C_{r3a}$ and $C_{w3b}$/$C_{r3b}$ addresses a fundamental incompatibility: liveness monitoring requires a counter that resets on every beat to catch per-beat stalls, while burst-length validation requires monotonic accumulation across all beats to detect premature, missing, or excess \texttt{WLAST}/\texttt{RLAST}. A single counter cannot serve both roles simultaneously. The dual-counter scheme therefore assigns each role to a dedicated counter, as reflected by the segmented and continuous bars in Figure~\ref{fig:plt_phases}: the liveness counter ($C_{w3a}$/$C_{r3a}$) resets on every data handshake, detecting per-beat stalls regardless of burst length; the beat counter ($C_{w3b}$/$C_{r3b}$) accumulates monotonically and is compared against \texttt{AWLEN}+1 or \texttt{ARLEN}+1 at \texttt{WLAST}/\texttt{RLAST}; a mismatch indicates a premature, missing, or excess \texttt{LAST} and triggers fault management.

The pre-enqueue watchdogs $C_{w0}$ and $C_{r0}$ are shared channel-level counters (Shared\textsuperscript{a}): one per channel, reset on each new \texttt{AW}/\texttt{AR} handshake. Sharing a single instance per channel is valid because AXI4 does not permit a second \texttt{AW}/\texttt{AR} handshake to complete before the first has been accepted. This ensures that only one request can be in the pre-enqueue phase at any time.

\textit{Prescaled Timebase.} All TMU variants share a prescaled timebase from a programmable divider (\texttt{PrescalerDiv}). Counters increment only on coarse ticks; event signals are sampled with sticky logic to prevent missed transitions between ticks. This reduces counter bit-width proportionally, trading detection resolution for area efficiency.

\begin{table}[t]
\centering
\caption{TMU register interface.}
\label{tab:tmu_registers}
\begin{tabular}{llm{5cm}}
\toprule
\textbf{Register} & \textbf{R/W} & \textbf{Description} \\
\midrule
budget[i]      & RW & Timeout threshold for phase counter $i$ \\
\addlinespace
prescaler\_div & RW & Prescaler divider; all counters increment on coarse tick \\
\addlinespace
latency[i]     & RO & Most recently observed latency for phase $i$ \\
\addlinespace
fault\_status  & RO & Latched first-fault metadata: fault type, transaction ID,
                      offending counter, direction, detection latency \\
\addlinespace
fault clear & WO & Write to acknowledge fault\rev{, lift cut phase,}
                            and re-arm monitor \\
reset req   & RO & Asserted on subordinate-induced fault; drives external reset controller \\
\bottomrule
\end{tabular}
\vspace{-4pt}
\end{table}

\subsubsection{Dequeue and Protocol Validation}
\label{sec:plt:dequeue}

To enforce per-ID in-order completion (Rules~W2/R2) while supporting out-of-order responses across IDs (Rules~W1/R0), the TMU always dequeues from the head of each ID's chain. On B handshake, \texttt{BID} triggers a parallel associative lookup over occupied ID Queue buckets. Upon match, the head Transaction Table slot is retrieved via the head pointer, and $C_{w3b}$ is checked for burst-length 
correctness against \texttt{AWLEN}+1. The slot is freed, head advances to \texttt{next}, and if the per-ID chain becomes empty, the bucket is released. The \texttt{RLAST} dequeue follows the same pattern via \texttt{RID} lookup, validating $C_{r3b}$ against \texttt{ARLEN}+1. Any mismatch, including an unmatched BID/RID, triggers fault management with full diagnostic metadata.

\subsubsection{Fault Management}
\label{sec:plt:fault}
The \textit{cut-and-drain} mechanism introduced in Section~\ref{sec:archi} is realized in PLT through the LQA structures. The \textit{cut} phase asserts sustained backpressure (\texttt{AWREADY}$\,{=}\,$0, \texttt{ARREADY}$\,{=}\,$0) to prevent new transactions from entering the affected subordinate path. The \textit{drain} phase operates in two stages: a \textit{fabrication arbiter} priority-selects pending transactions by iterating over occupied ID Queue buckets, and a \textit{response fabricator} synthesizes protocol-compliant completions, one B response (\texttt{BRESP}$\,{=}\,$\texttt{SLVERR}) per pending write and \rev{the remaining read data beats up to \texttt{ARLEN+1} with \texttt{RRESP}$\,{=}\,$\texttt{SLVERR}} per pending read, each carrying the correct remapped ID. A response multiplexer selects between subordinate-generated and TMU-fabricated responses, ensuring all issued requests receive valid completions to prevent deadlocks. Each LQA entry is freed as soon as its fabricated response is accepted by the manager, progressively clearing structures until all transactions are retired.

\rev{In PLT and CLT, completing the burst is required rather than optional: AXI4 defines no early termination, so an error response must still return the full \texttt{ARLEN+1} transfers, with \texttt{RLAST} on the final beat. ILT, which stores no per-transaction burst length, instead retires each pending read with a single beat asserting \texttt{RLAST}. The write path carries no such obligation: a single B response retires a transaction irrespective of the W beats transferred, and Rule~W0 leaves at most one residual data phase for the TMU to absorb. Section~\ref{sec:drain} bounds the resulting drain time for all three variants.}

The \textit{Fault Aggregation and Logging} consolidates timeout and protocol-violation signals into a single first-fault latch. On the first fault, the latch captures the faulting transaction identifier, access direction (read or write), the faulty transaction phase, the fault detection latency (in cycles from fault onset to interrupt assertion), and fault-type indicators that distinguish timeouts from protocol violations. This latch holds until software clears the \texttt{fault\_clear} register. Because the cut-and-drain intentionally generates error responses that may trigger secondary protocol events, the aggregator masks all subsequent faults to preserve root-cause integrity.

\begin{figure}[htbp]
\centerline{\includegraphics[width=0.52\textwidth]{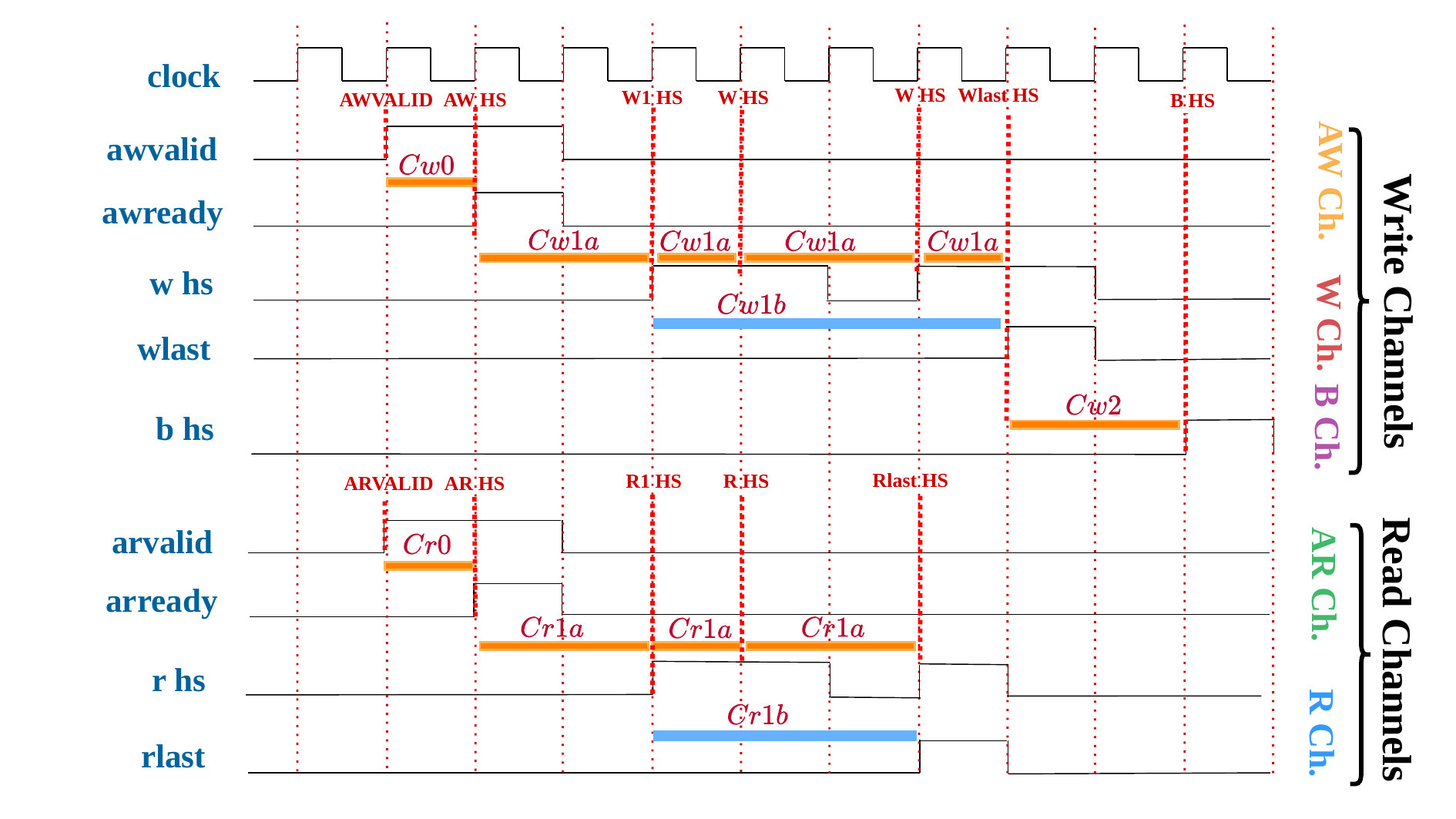}}
\caption{Channel-level counter allocation in CLT write and read paths.}
\label{fig:clt_stages}
\vspace{-8pt}
\end{figure}

\subsubsection{Programming Model}
\label{sec:plt:progmodel}

The TMU is managed via the register interface detailed in Table~\ref{tab:tmu_registers}. Timeout budgets and the prescaler are runtime-configurable. The \texttt{latency[i]} registers record the most recently observed latency for each phase counter $i$, capturing per-phase timing under both nominal and faulty conditions and providing a measurement baseline for \texttt{budget[i]} threshold tuning. Budgets can be configured through two complementary workflows: (a) online, by inspecting \texttt{latency[i]} under representative traffic and setting \texttt{budget[i]} margins above observed peaks to minimize false-positive rates; and (b) offline, by deriving upper latency bounds from static workload analysis of known applications and using these bounds to set \texttt{budget[i]} 
thresholds that guarantee fault detection within the required timing constraints. \rev{Upon fault detection, the TMU asserts \texttt{irq} and, for
subordinate-induced faults, \texttt{reset\_req} to the platform reset controller, whose sequencing is device-specific. Cut-and-drain and the reset request are autonomous, so interconnect liveness does not depend on interrupt latency. The interrupt service routine (ISR) reads \texttt{fault\_status} (Table~\ref{tab:tmu_registers}), which holds the first-fault snapshot until acknowledged, and writes \texttt{fault\_clear} to lift the cut phase and re-arm the monitor. Because AXI4 delegates error signalling to the originating manager via \texttt{BRESP}/\texttt{RRESP}, the TMU reports the fault but leaves the recovery policy, whether to reissue, skip, or escalate, to the manager and its driver.}

\vspace{-5pt}
\subsection{Channel-Level Tracking (CLT)}
\label{sec:clt}

While PLT provides maximum observability, its per-phase counter replication imposes non-negligible area overhead for less critical or lower-bandwidth subordinates in mixed-criticality SoCs. CLT trades fine-grained phase localization for substantially reduced counter complexity while preserving deadlock prevention and protocol violation detection (Table~\ref{tab:tmu_comparison}).

\subsubsection{Channel Counter Organization}
\label{sec:clt:counters}

CLT retains PLT's foundational architecture \rev{(the ID Remapper, LQA, and
pre-enqueue watchdogs of Section~\ref{sec:plt})} but coalesces per-phase
counters into channel-level liveness monitors. Where PLT requires twelve counters (six timeout plus one beat counter for writes, four timeout counters plus one beat counter for reads), CLT uses only seven: three timeout counters ($C_{w0}$, $C_{w1a}$, $C_{w2}$) plus one beat counter ($C_{w1b}$) on the write path, and two timeout counters ($C_{r0}$, $C_{r1a}$) plus one beat counter ($C_{r1b}$) on the read path (Figure~\ref{fig:clt_stages}). Individual \texttt{VALID}/\texttt{READY} pairs are abstracted into unified handshake events (\texttt{W\_HS}, \texttt{B\_HS}, \texttt{R\_HS}) to monitor channel-level progress rather than signal-level stalls. CLT detects channel stalls, missing responses, burst-length mismatches, and \texttt{BID}/\texttt{RID} violations, but cannot localize faults to specific sub-phases within channels. \rev{A timeout on the data or response channels identifies a stalled channel without attributing manager versus subordinate responsibility; \texttt{reset\_req} is therefore issued on address-phase faults alone.}

\begin{figure}[htbp]
\centerline{\includegraphics[width=0.52\textwidth]{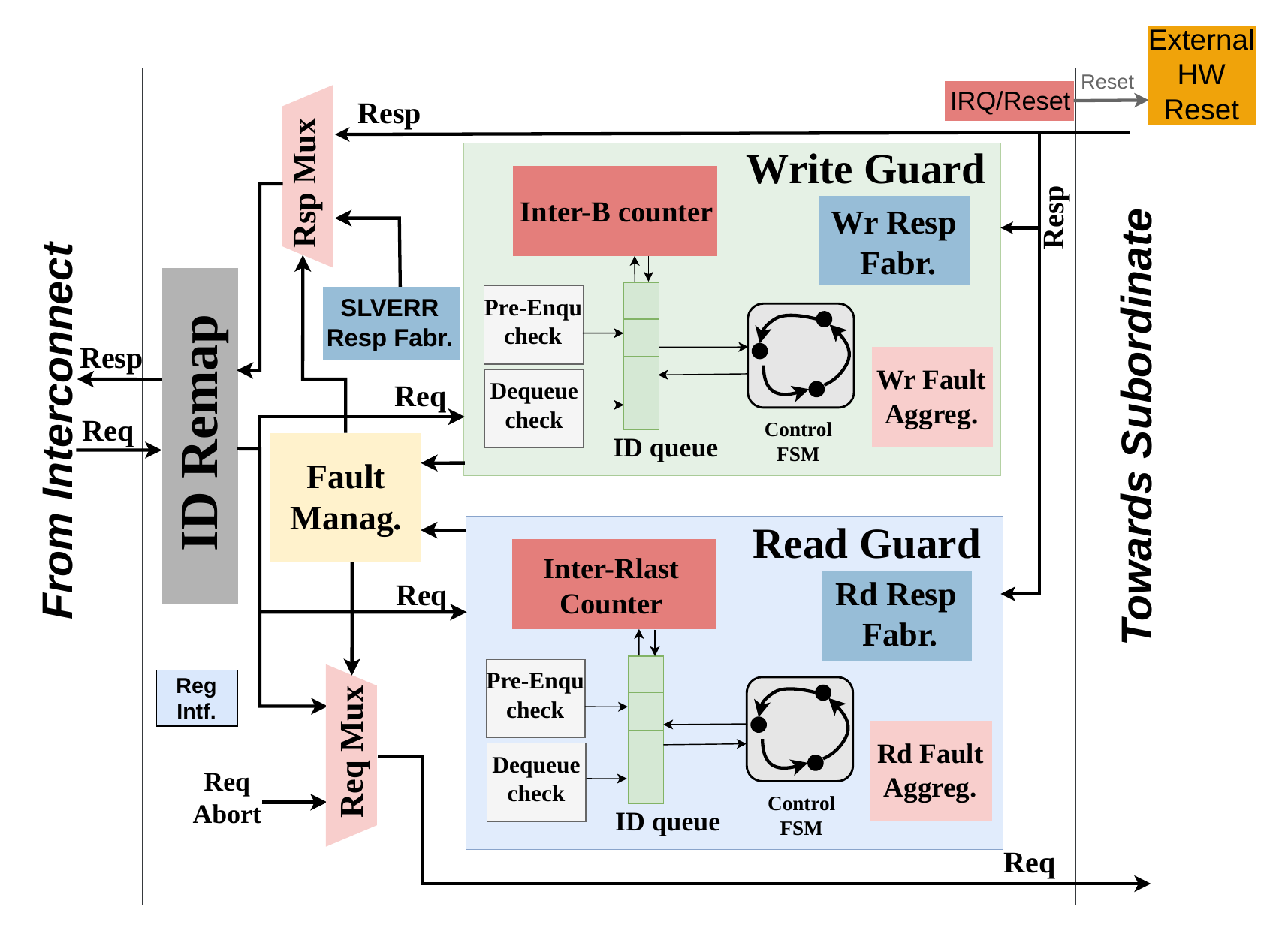}}
\vspace{-8pt}
\caption{ILT architecture. The Transaction Table and W-Ownership FIFO are eliminated; only a simplified ID Queue tracking outstanding transaction counts per ID is retained.}
\label{fig:ilt_top}
\vspace{-6pt}
\end{figure}

\subsubsection{Write Path Monitoring}
$C_{w0}$ is the pre-enqueue \texttt{AW} watchdog, identical to PLT (Section~\ref{sec:plt:preenqueue}). Because Rule~W0 
serializes all W-data through a single channel, one shared liveness counter suffices: $C_{w1a}$ is assigned to the W-FIFO 
head transaction, activates whenever a new transaction is promoted to head, increments each cycle in the absence of a W 
handshake, and resets on each W beat, jointly covering first-beat and all inter-beat stalls and replacing PLT's separate $C_{w1}$, $C_{w2}$, and $C_{w3a}$. The paired beat counter $C_{w1b}$ accumulates on each W handshake and is compared against \texttt{AWLEN}+1 at \texttt{WLAST}; after \texttt{WLAST}, $C_{w2}$ activates per-ID in the corresponding ID Queue bucket to monitor B-channel response arrival; because Rule~W2 guarantees in-order B completion within each ID, the same counter instance is reused temporally for subsequent transactions as they become the per-ID chain head.

\subsubsection{Read Path Monitoring}
$C_{r0}$ is the pre-enqueue \texttt{AR} watchdog. Per-ID allocation for R-channel counters is mandatory because Rule~R1 interleaving would cause a shared counter to be prematurely reset by beats from other IDs. $C_{r1a}$ therefore activates per-ID in the corresponding ID Queue bucket upon address-handshake completion, resetting on each R handshake for that ID. This merges PLT's $C_{r1}$, $C_{r2}$, and $C_{r3a}$ into a single channel liveness counter. The paired beat counter $C_{r1b}$ accumulates per-ID and is compared against \texttt{ARLEN}+1 at \texttt{RLAST}; no separate response-phase counter is required, since \texttt{RLAST} serves as both the final data beat and transaction completion event. The same $C_{r1a}$/$C_{r1b}$ pair is reused temporally across successive transactions within each ID per Rule~R2.

CLT exposes the interface of (Table~\ref{tab:tmu_registers}) with \texttt{budget[i]} and \texttt{latency[i]} reduced to seven indices, reflecting CLT's seven-counter structure and \texttt{fault\_status} provides coarser diagnostic statistics, reporting at channel granularity.

\vspace{-5pt}
\subsection{ID-Level Tracking (ILT)}
\label{sec:ilt}

\begin{figure}[htbp]
\centerline{\includegraphics[width=0.52\textwidth]{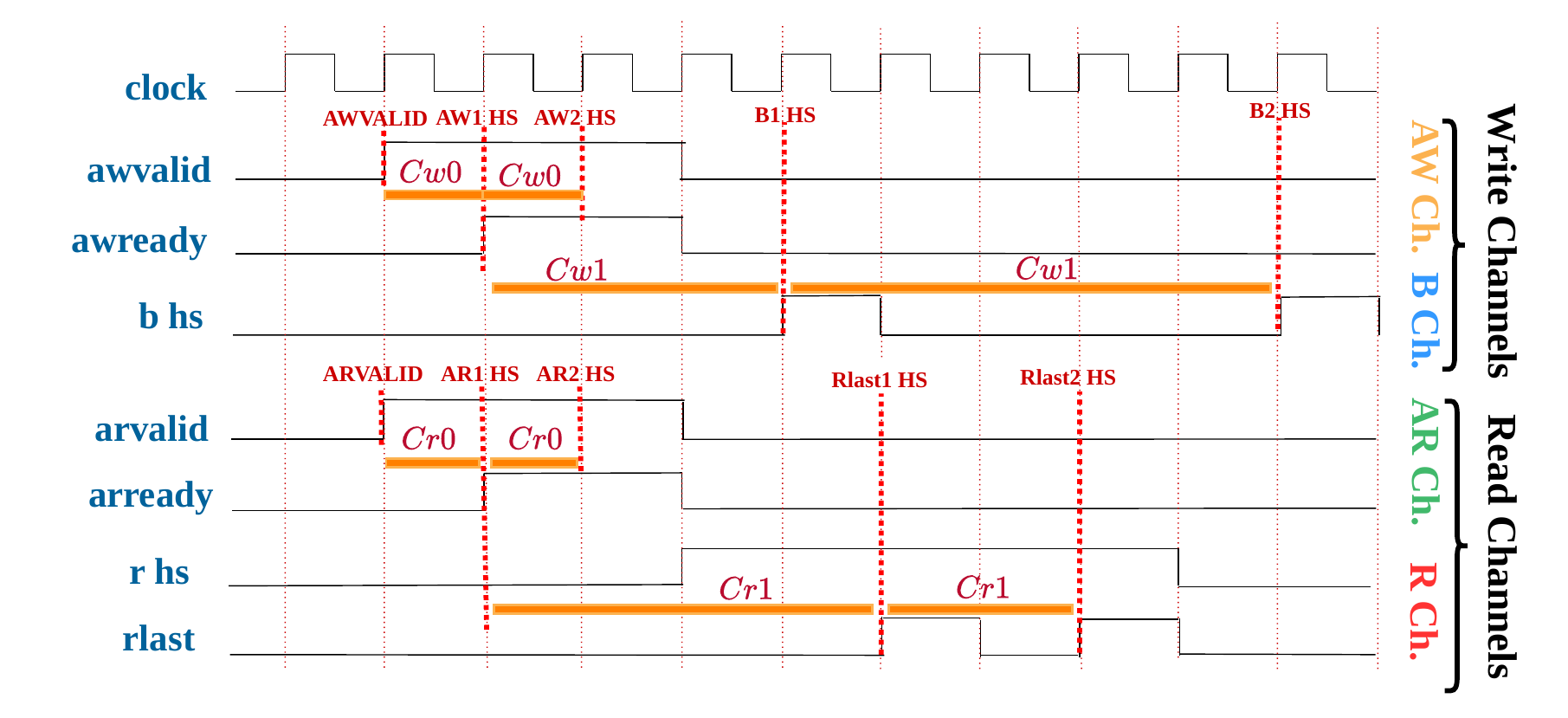}}
\vspace{-10pt}
\caption{Counter allocation in ILT: pre-enqueue watchdogs ($C_{\textsubscript{w0}}$, $C_{\textsubscript{r0}}$) and per-ID completion timeouts ($C_{\textsubscript{w1}}$, $C_{\textsubscript{r1}}$).}
\label{fig:ilt_stage}
\vspace{-4pt}
\end{figure}

\subsubsection{Simplified Architecture}
\label{sec:ilt:arch}

While CLT reduces timeout counters relative to PLT, both variants rely on the LQA to store per-transaction metadata such as burst lengths for beat validation and linked-list pointers for ordering enforcement, resulting in $\mathcal{O}(N_{ID} \times D)$ area scaling (Table~\ref{tab:tmu_comparison}). For highly area-constrained subordinates with deep transaction 
concurrency, this depth-dependent storage remains prohibitive. ILT eliminates this dependency entirely by replacing per-transaction tracking with per-ID completion monitoring, reducing area complexity to $\mathcal{O}(N_{ID})$ at the cost of detection granularity and latency.

Figure~\ref{fig:ilt_top} shows ILT's architecture: the Transaction Table and
W-Ownership FIFO \rev{of Section~\ref{sec:plt}} are eliminated, leaving only a
simplified ID Queue. Each entry contains four fields: remapped transaction ID, \texttt{num\_txn} (which dynamically tracks the ID's real-time outstanding depth by incrementing upon an address handshake (\texttt{AW}/\texttt{AR}) and decrementing upon a completion handshake (\texttt{B}/\texttt{RLAST})), a per-ID timeout counter ($C_{w1}$ for writes, $C_{r1}$ for reads), and a free bit.

This reduction to a single timeout counter per ID is enabled by two complementary arguments. First, AXI4's in-order completion guarantee within the same ID (Rules~W2/R2) means that monitoring only the completion boundary is sufficient to guarantee forward progress for the entire queued chain. Second, any stall in an intermediate phase, whether at the address, data, or response channel, will propagate to the completion boundary: a manager or subordinate that stalls mid-transaction cannot issue \texttt{B}/\texttt{RLAST}, so $C_{w1}$/$C_{r1}$ will always fire regardless of where the deadlock originates. \texttt{num\_txn} provides the minimal concurrency tracking needed to determine when an ID has outstanding transactions requiring supervision. The diagnostic consequence is that fault reporting identifies \emph{which} ID stalled but cannot isolate the specific transaction, burst beat, or protocol phase responsible.

Figure~\ref{fig:ilt_stage} shows ILT's four counters (vs.\ PLT's twelve, CLT's seven): $C_{w0}$/$C_{r0}$ pre-enqueue watchdogs and $C_{w1}$/$C_{r1}$ per-ID completion timeouts.

\subsubsection{Write Path Monitoring}
\label{sec:ilt:write}

$C_{w0}$ is the pre-enqueue \texttt{AW} watchdog, identical to PLT and CLT (Section~\ref{sec:plt:preenqueue}). Once enqueued, $C_{w1}$ acts as the per-ID completion timeout: for the first transaction of an ID it activates on \texttt{AW} handshake, tracking latency to the first B response; for subsequent pipelined transactions it operates as an Inter-B counter, resetting on each B handshake. The counter remains active while \texttt{num\_txn}~$>~0$. Because $C_{w1}$ is allocated strictly per-ID, concurrent progress on other IDs cannot reset a stalled ID's timer, ensuring independent fault detection across IDs. Burst-length validation and \texttt{WLAST} protocol checking are not supported.

\subsubsection{Read Path Monitoring}
\label{sec:ilt:read}
Read monitoring is symmetric. $C_{r0}$ is the pre-enqueue \texttt{AR} watchdog. $C_{r1}$ activates on \texttt{AR} handshake for the first transaction of an ID and thereafter operates as an Inter-\texttt{RLAST} counter, resetting on each \texttt{RLAST} while \texttt{num\_txn} $>0$. Intermediate R-channel stalls are detected only if they delay \texttt{RLAST} beyond the $C_{r1}$ budget. \texttt{ARLEN} validation and premature or excess \texttt{RLAST} detection are not supported.

ILT exposes the register interface (Table~\ref{tab:tmu_registers}) unchanged, with \texttt{budget[i]} and \texttt{latency[i]} reduced to four indices and \texttt{fault\_status} reporting at ID granularity.
\vspace{-5pt}
\section{Worst-Case Detection Time Analysis}
\label{sec:wcdt}

The Worst-Case Detection Time (WCDT) characterizes the theoretical upper bound on TMU fault-detection latency. 
It is governed by architectural granularity (phase-level, channel-level, or ID-level) and by the configured timeout budgets through control registers at system startup or runtime.

Let $t_\text{fault}$ denote the cycle at which a fault induces an error in the target manager or subordinate, $t_\text{start}$ the cycle at which the affected transaction enters the AXI protocol domain, and $t_\text{detect}$ the cycle at which the TMU flags the violation and initiates recovery. Since faults may remain dormant in internal logic before manifesting at the bus interface, $t_\text{fault}$ is often unobservable to interconnect monitors. Therefore, the effective WCDT is defined as the fault exposure window:
\begin{equation}
    \text{WCDT} = t_\text{detect} - t_\text{start}
    \label{eq:wcdt}
\end{equation}
where the observable entry point $t_\text{start}$ corresponds to transaction initiation (\texttt{AWVALID} or \texttt{ARVALID} assertion). This window represents the maximum duration a faulty transaction remains active, potentially stalling the bus or consuming tracking resources before being suppressed by the TMU.

\vspace{-12pt}
\subsection{Phase-Level Tracking (PLT)}

PLT monitors each protocol phase independently, so detection occurs immediately upon expiration of the specific failing phase's budget. For a fault occurring in phase $i \in \Phi$, where $\Phi$ is the ordered set of write or read phases:
\begin{equation}
    \text{WCDT}_\text{PLT}^{(i)} = \sum_{j=0}^{i-1} \tau_j 
    + \text{Th}_i
     \label{eq:wcdt_plt}
\end{equation}
where $\tau_j$ is the nominal completion time of antecedent phase $j$ and $\text{Th}_i$ is the timeout threshold of the failing phase $i$. The overall WCDT for a transaction is the maximum across all possible failing phases:
\vspace{-5pt}
\begin{equation}
    \text{WCDT}_\text{PLT} = \max_{i \in \Phi}
    \left( \sum_{j=0}^{i-1} \tau_j + \text{Th}_i \right)
\end{equation}
Because $\text{Th}_i$ is tightly bounded to the physical latency requirement of phase $i$ alone, PLT yields the lowest fault exposure window among the three architectures.

\vspace{-12pt}
\subsection{Channel-Level Tracking (CLT)}

CLT collapses multiple fine-grained phases into unified channel-level monitors. The WCDT formula remains structurally similar over the reduced set of channel monitoring points $K$:
\vspace{-5pt}
\begin{equation}
    \text{WCDT}_\text{CLT} = \max_{k \in K}
    \left( \sum_{j=0}^{k-1} \tau_j + 
    \text{Th}_k^\text{CLT} \right)
\end{equation}
The critical distinction lies in $\text{Th}_k^\text{CLT}$: because a single counter must accommodate the worst-case timing of all PLT phases collapsed into channel monitor $k$, the 
unified budget must satisfy $\text{Th}_k^\text{CLT} \geq \max_{i \in k}\,\text{Th}_i$, where $i$ indexes the PLT phases subsumed by $k$. This inflates the unified budget above any individual PLT threshold, thus increasing the WCDT relative to PLT.

\vspace{-12pt}
\subsection{ID-Level Tracking (ILT)}

ILT monitors only address handshakes and final completion events (\texttt{B}/\texttt{RLAST}), with no visibility into intermediate protocol phases. The WCDT is governed by two budgets:
\begin{equation}
    \text{WCDT}_\text{ILT} = \max\left(
    \text{Th}_\text{Addr},\, \text{Th}_\text{End}^\text{ILT}
    \right)
    \label{eq:wcdt_ilt}
\end{equation}
\vspace{-13pt}

where $\text{Th}_\text{Addr}$ is the pre-enqueue address watchdog and $\text{Th}_\text{End}^\text{ILT}$ is the completion timeout (Inter-B or Inter-\texttt{RLAST}). When an ID transitions from idle to active, the completion counter activates at the address handshake and must span the full address-to-completion latency, covering the address, data, and 
response phases in a single budget. Because no intermediate phase is observable, $\text{Th}_\text{End}^\text{ILT}$ is necessarily larger than any individual PLT phase threshold, yielding the largest WCDT among the three architectures:
\begin{equation}
    \text{WCDT}_\text{PLT} \leq \text{WCDT}_\text{CLT}
    \leq \text{WCDT}_\text{ILT}
\end{equation}

\vspace{-12pt}
\subsection{Cut-and-Drain Time}
\label{sec:drain}

The WCDT bounds detection; restoring liveness additionally requires
draining the transactions outstanding when the cut begins. Let $N_w$ and
$N_r$ denote the outstanding writes and reads at that instant, and
$L_{\max} = \mathtt{AxLEN}_{\max} + 1$ the maximum burst length. The cut
admits no further requests, so the drain set is fixed once recovery
starts, and is bounded by construction: \texttt{MaxOutstdTxns} per
direction in PLT and CLT, and $\mathtt{MaxUniqIDs} \times
\mathtt{TxnPerUniqID}$ in ILT, whose per-ID \texttt{num\_txn} counters
serve the same role.

The two paths drain concurrently and their costs differ. Rule~W0
serializes W-data by AW-issue order, so at most one write occupies its
data phase at the cut; absorbing its residual beats costs at most
$L_{\max}$ cycles, after which one fabricated B response with
$\mathtt{BRESP}=\mathtt{SLVERR}$ retires each pending write, irrespective
of how many W beats were transferred. Rule~R1 permits reads with distinct IDs to be mid-burst simultaneously, and each must be completed to its full \texttt{ARLEN+1} beats (Section~\ref{sec:plt}). With one fabricated beat
accepted per cycle on each channel,
\begin{equation}
\begin{split}
T_{\mathrm{drain}} &\leq \max(L_{\max} + N_w,\; N_r \cdot L_{\max}) \\
                   &\leq \max(L_{\max} + M,\; M \cdot L_{\max}),
\end{split}
\end{equation}
where $M = \mathtt{MaxOutstdTxns}$. Beyond single-beat bursts the read
term dominates, giving $T_{\mathrm{drain}} \leq M \cdot L_{\max}$: drain
time is linear in capacity on both paths and multiplicative in burst
length on the read path. ILT is the exception: storing no per-transaction
burst length, it retires each pending read with a single beat asserting
\texttt{RLAST}, giving $T_{\mathrm{drain}} \leq \max(L_{\max} + N_w,
N_r)$, independent of burst length. This requires managers that treat \texttt{RLAST} as the transaction boundary rather than counting beats against \texttt{ARLEN}, a deployment condition on ILT and a further consequence of the reduced state behind its $\mathcal{O}(N_{ID})$ scaling.

\begin{figure}[htbp]
\centering
\hspace{-1cm}
\includegraphics[width=0.3\textwidth, height=0.2\textwidth]{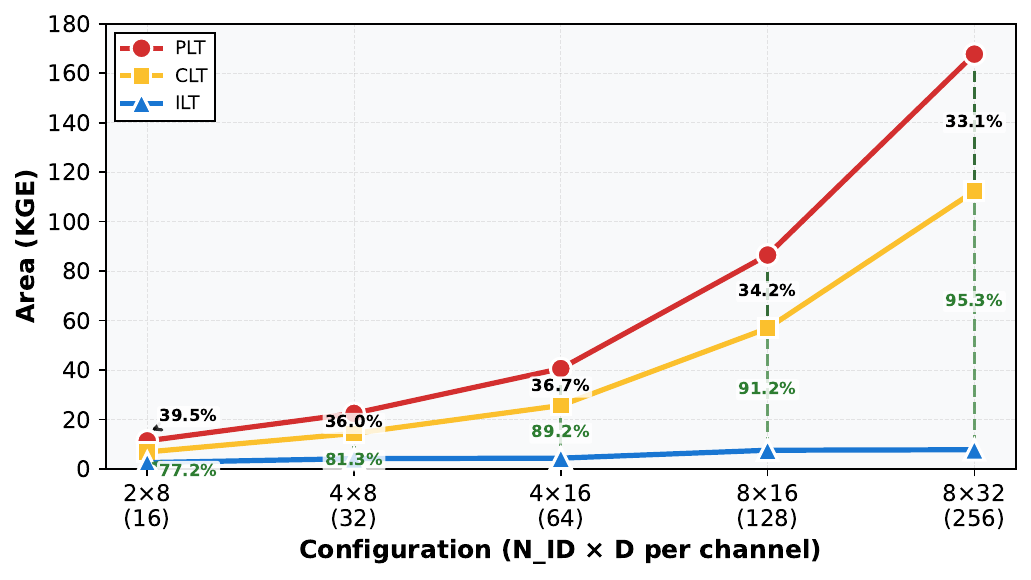}
\caption{TMU area scaling with annotated reduction percentages.}
\label{fig:area_comp}
\vspace{-6pt}
\end{figure}

Liveness is therefore restored within
\begin{equation}
T_{\mathrm{liveness}} = \mathrm{WCDT} + T_{\mathrm{drain}},
\end{equation}
with both terms fixed at design time. The bound is independent of the subordinate's state, since the fabricator sources every response locally: a completely unresponsive subordinate drains no more slowly than a partially functional one. It covers the hardware path only. Software recovery, comprising interrupt delivery, the ISR, reset sequencing, and driver reinitialization before \texttt{fault\_clear} (Section~\ref{sec:plt}), depends on the platform and the device and is not bounded here.
\section{Experimental Results}
\label{sec:results}

We evaluate PLT, CLT, and ILT in terms of silicon area, timing closure, power, and fault-detection coverage and latency.

\vspace{-12pt}
\subsection{Area Overhead and Scaling Characteristics}

\begin{figure*}[htbp]
\centering
\includegraphics[width=1\textwidth,height=0.31\textwidth]{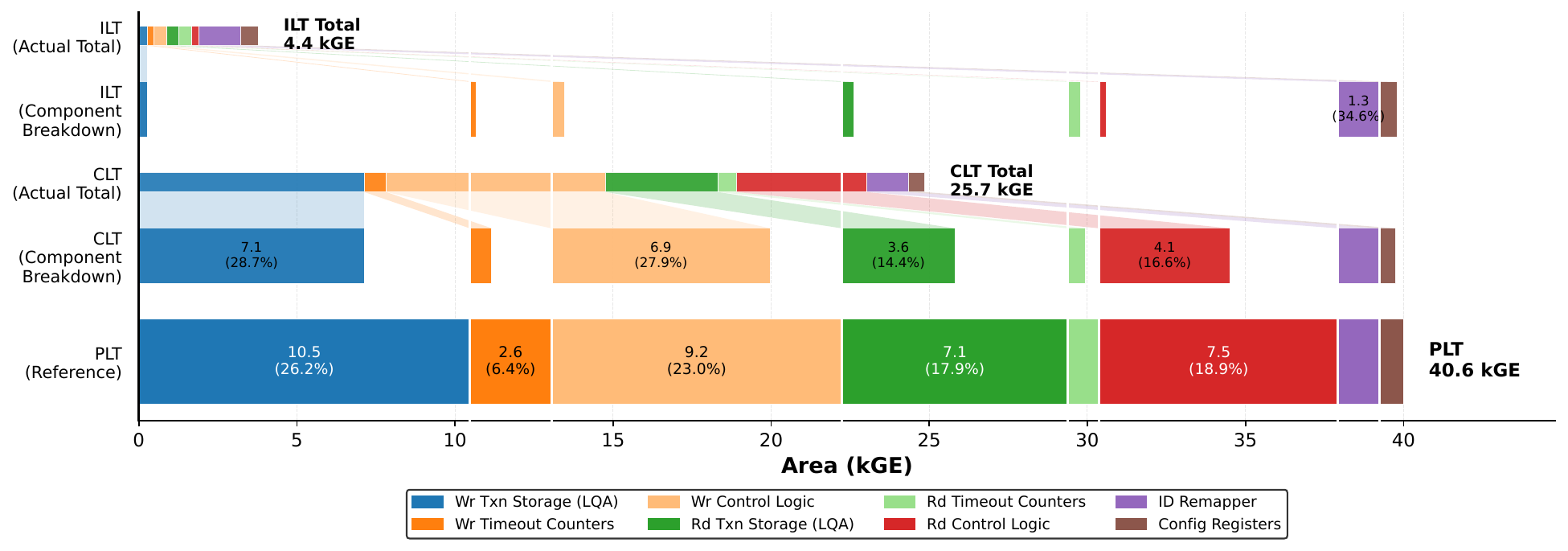}
\vspace{-10pt}
\caption{Component-level area breakdown for $4 \times 16$ configuration.}
\label{fig:breakdown}
\vspace{-15pt}
\end{figure*}

All TMU variants were synthesized using GlobalFoundries' 12nm FinFET process (GF12) targeting a 1~GHz clock, representative of a typical operating frequency of mixed-criticality SoC interconnects~\cite{b26,b_realm}. We evaluate five configurations spanning 16 to 256 total outstanding transactions \textit{per direction} (write or read), defined by $N_{ID} \times D$, where $N_{ID}$ denotes the number of unique AXI transaction IDs and $D$ the maximum outstanding transactions per ID. Since AXI provides independent write and read channels, each configuration supports twice this capacity system-wide (e.g., $4 \times 16$ enables 128 total outstanding transactions). These cover deployment scenarios from resource-constrained embedded peripherals ($2 \times 8$) to high-performance memory controllers targeting high-end automotive and aerospace deployments ($8 \times 32$), covering the practical concurrency range of state-of-the-art AXI4-based mixed-criticality platforms~\cite{b10,b_realm}.

Figure~\ref{fig:area_comp} illustrates the area scaling characteristics of
the three architectures. PLT and CLT area grow proportionally to total
monitoring capacity ($\mathcal{O}(N_{ID} \times D)$) due to their
per-transaction state dependencies, PLT scaling from 11.4\,kGE at
$2\times8$ to 167.7\,kGE at $8\times32$ and CLT from 6.9\,kGE to
112.3\,kGE, a stable 33.1\% to 39.5\% reduction over PLT across all
configurations. ILT achieves superior scalability
($\mathcal{O}(N_{ID})$) by decoupling its area from $D$ entirely, growing
only $3.0\times$ (2.6\,kGE to 7.8\,kGE) for the same $16\times$ capacity
increase and saving between 77.2\% ($2\times8$) and 95.3\% ($8\times32$)
over PLT. The cost is a $3.7\times$ higher median detection latency at
$4\times16$ (Section~\ref{sec:wcdt_eval}).

\begin{figure}[htbp]
\centering
\hspace{-1cm}
\includegraphics[width=0.45\textwidth, height=0.18\textwidth]{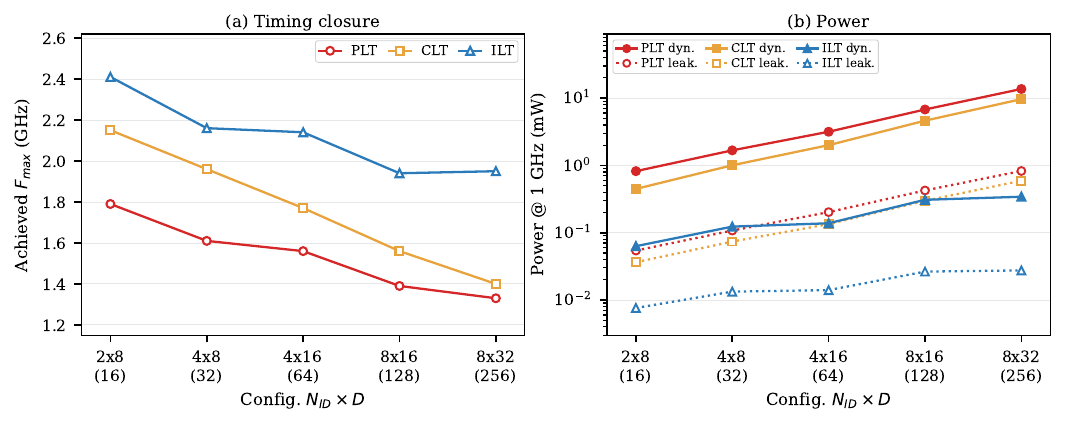}
\vspace{-8pt}
\caption{Achieved $F_{\max}$ (a), and leakage and dynamic power (b), across
configurations.}
\label{fig:fmax_power}
\vspace{-6pt}
\end{figure}

Figure~\ref{fig:breakdown} decomposes the TMU area at the $4 \times 16$ configuration, showing how architectural abstraction redistributes silicon cost across components. PLT (40.6\,kGE) dedicates 10.5\,kGE to write transaction storage and 7.1\,kGE to read storage for per-phase coverage. CLT (25.7\,kGE) achieves its 36.7\% reduction by consolidating per-phase transaction state into unified channel watchers, dropping write storage to 7.1\,kGE and read to 3.6\,kGE. ILT (4.4\,kGE) achieves its area reduction by monitoring only 8 transaction IDs (4 per direction) rather than 128 concurrent transactions, collapsing per-transaction logic to negligible area. As $D$ grows from 8 to 32, ILT area remains dominated by the fixed $N_{ID}$ tracking structure, directly explaining Figure~\ref{fig:area_comp}'s sub-linear curve.

\rev{Timing Closure and Power. Fig.~\ref{fig:fmax_power} reports achieved $F_{\max}$, leakage, and dynamic power for all configurations. \rev{All 15 configurations close the 1\,GHz target with positive slack; the minimum achieved $F_{\max}$ is 1.33\,GHz (PLT at $8\times32$), so no
configuration is timing-critical at the reported operating point.}
Power is reported at the 1\,GHz iso-frequency point, with dynamic power
back-annotated from the in-system iDMA traffic of Section~\ref{sec:insystem}.}

\vspace{-10pt}
\subsection{In-System Evaluation}
\label{sec:insystem}

\begin{figure}[htbp]
\centering
\includegraphics[width=0.48\textwidth]{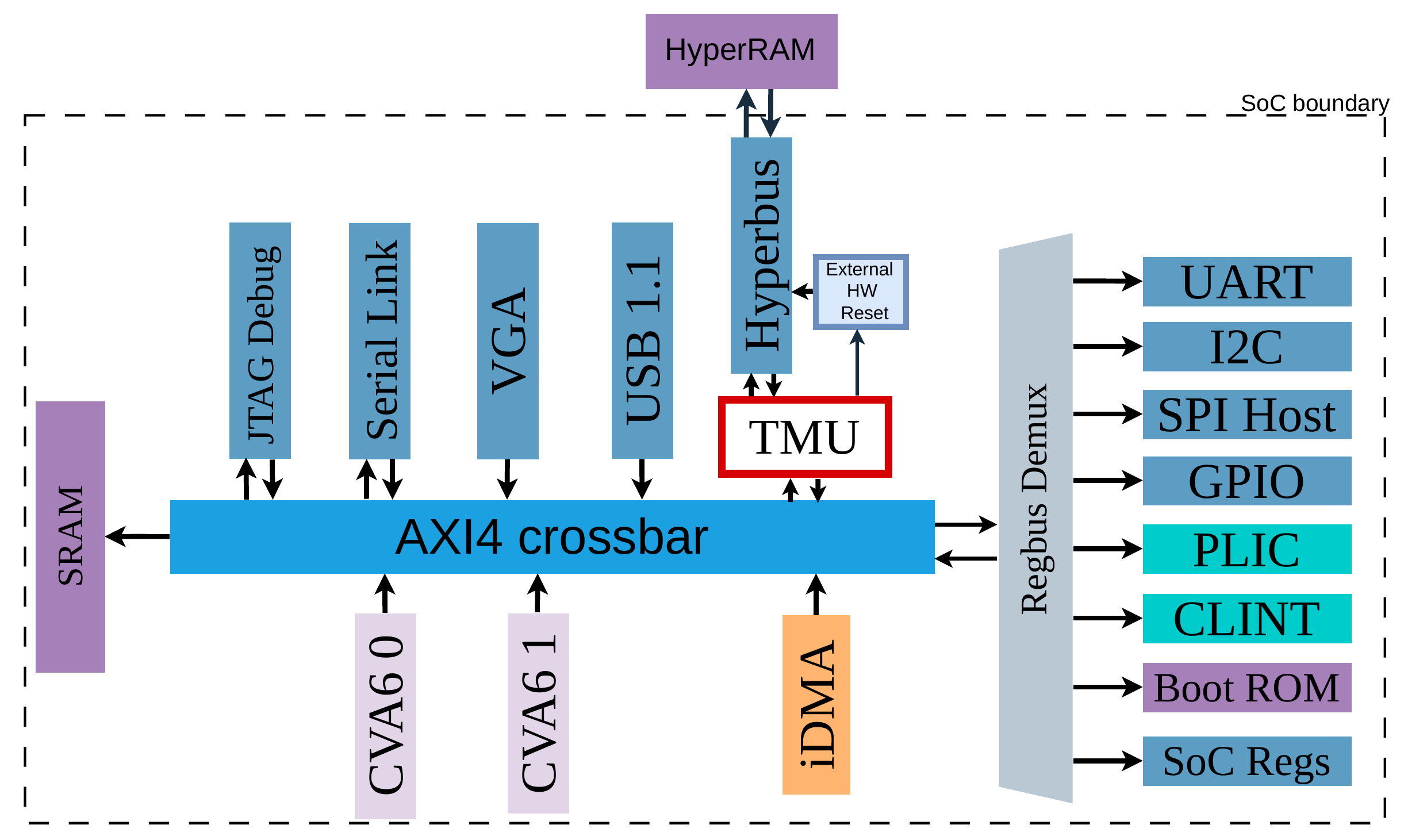}
\caption{TMU integration within the Cheshire SoC, positioned between the AXI4 crossbar and HyperBus controller to monitor iDMA-driven traffic between on-chip SRAM and external HyperRAM.}
\label{fig:chs}
\vspace{-10pt}
\end{figure}

All TMU variants were integrated into the Cheshire SoC~\cite{b10}, an open-source heterogeneous 64-bit RISC-V platform built around the CVA6 core~\cite{zaruba2019,cva6virt}. The TMU is instantiated between the main AXI4 crossbar and the HyperBus controller (Figure~\ref{fig:chs}), monitoring all AXI4 transactions to off-chip HyperRAM in real time. The iDMA engine performs continuous transfers between internal SRAM and external HyperRAM, emulating the double-buffering data movement patterns typical of DSP and AI computing pipelines on mixed-criticality SoCs, and generating sustained bidirectional AXI4 write and read traffic.

\subsubsection{Fault Injection Campaign}

To quantify the TMU's diagnostic coverage and validate the theoretical WCDT bounds (Section~\ref{sec:wcdt}) under transient faults, we conduct fault injection campaigns using Synopsys VC Z01X~\cite{zoix}, a fault simulation platform for functional safety analysis compliant with ISO~26262~\cite{iso26262} and IEC~61508~\cite{iec61508} industrial standards, providing evidence-based diagnostic coverage measurement in conjunction with Synopsys VCS. Z01X automates fault list generation, injection scheduling, and outcome classification, providing standards-compliant diagnostic coverage metrics applicable to safety integrity level assessment.

Faults are injected into two IP cores separately to evaluate how each TMU architecture detects manager- and subordinate-induced violations and at what latency: the iDMA controller (AXI manager) and the HyperBus memory controller (AXI subordinate). We inject 100k faults per fault class per target IP and per TMU variant, yielding 12 independent campaigns in total, targeting sequential state elements (FLOP, modeling Single Event Upsets (SEUs)) and module port boundaries (PORT, modeling Single Event Transients (SETs)). Each scenario comprises an independent simulation with a single fault injected at a randomly selected cycle, executing 1,024 transactions (512 writes, 512 reads) as background traffic to ensure uniform temporal coverage. Fault outcomes are classified as: (1)~\textit{Masked (AM)}: no observable system-level effect; (2)~\textit{Detectable Faults (BA)}: AXI specification violations and stalls detectable by the TMU; and (3)~\textit{Undetectable Faults (EA)}: \rev{faults not detectable by the variant under evaluation; Section~\ref{sec:faultmodel} bounds the maximum detectable class, and Table~\ref{tab:tmu_comparison} the per-variant scope}. Our analysis focuses on BA faults as the safety-critical class the TMU is designed to detect.

\subsubsection{Comparative Fault Detection Coverage}

\begin{table}[t]
\centering
\caption{Fault Classification Results: Comparison of Detection Rates.}
\label{tab:fault_classification}
\begin{tabular}{llccc}
\toprule
\textbf{Injection Point} & \textbf{Arch} & \textbf{AM (\%)} & \textbf{BA (\%)} & \textbf{EA (\%)} \\
\midrule
\multirow{3}{*}{iDMA Flop}
& PLT & 94.92 & 1.65 & 3.43 \\
& CLT & 94.92 & 1.65 & 3.43 \\
& ILT & 94.92 & 1.62 & 3.46 \\
\midrule
\multirow{3}{*}{iDMA Port}
& PLT & 80.98 & 3.49 & 15.53 \\
& CLT & 80.98 & 3.43 & 15.59 \\
& ILT & 80.98 & 3.26 & 15.76 \\
\midrule
\multirow{3}{*}{Hyper Flop}
& PLT & 86.19 & 4.01 & 9.80 \\
& CLT & 86.19 & 3.98 & 9.83 \\
& ILT & 86.19 & 3.74 & 10.07 \\
\midrule
\multirow{3}{*}{Hyper Port}
& PLT & 85.04 & 5.38 & 9.58 \\
& CLT & 85.04 & 5.35 & 9.61 \\
& ILT & 85.04 & 4.97 & 9.99 \\
\bottomrule
\end{tabular}
\vspace{-8pt} 
\end{table}

Table~\ref{tab:fault_classification} reveals two key findings. First, the masked fault ratio (AM) is identical across all three architectures at every injection point, confirming that 
TMU granularity does not influence whether a fault manifests at the AXI interface; faults that leave bus signals unperturbed are inherently unobservable to any protocol-level monitor \rev{(Section~\ref{sec:faultmodel})}. Second, the aggregate non-masked fraction (BA + EA) is constant per
injection point, but its split shifts with granularity: ILT's
endpoint-only monitoring reclassifies as undetectable (EA) a small subset
that PLT and CLT catch through finer-grained supervision. The 0.41\% shift
at the HyperBus port represents intermediate faults such as missing data
beats within a burst that do not stall the address-to-completion boundary
and therefore fall outside ILT's observability window.

\subsubsection{Deterministic Detection Behavior}
\label{sec:wcdt_eval}

To empirically validate the theoretical WCDT models (Section~\ref{sec:wcdt}), we configured a controlled campaign with uniform 32-beat bursts (\texttt{AxLEN}=\texttt{0x1F}) and alternating iDMA transfers. Fixing the burst length isolates architectural detection latency from workload variance. As a first step, we characterized per-phase transaction latencies under fault-free operation on the target HyperBus subordinate, obtaining the nominal latency measurements $\tau$ reported in Table~\ref{tab:wcdt_eval}. Timeout thresholds $\text{Th}$ were then set by adding a margin above the observed nominal values, following the offline static characterization workflow described in Section~\ref{sec:plt:progmodel}. \rev{No false detections were observed in any campaign, since by
construction no fault-free interval can reach its threshold. Each
threshold enters \eqref{eq:wcdt_plt}--\eqref{eq:wcdt_ilt} additively,
so margin traded for false-positive immunity inflates the detection bound
by the same amount; deriving budgets that jointly meet application
deadlines and bound false-positive probability is a schedulability
problem outside the scope of this paper. The trade-off is local rather
than system-wide: every counter resets on its protocol event and the
measurement window opens only at the monitored interface
\eqref{eq:wcdt}, so neither burst length, outstanding count, nor
upstream queuing enlarges a supervised interval. ILT is the exception,
its completion timeouts spanning transaction boundaries and widening
under cross-ID interleaving, as its larger budgets in
Table~\ref{tab:wcdt_eval} reflect. Tightness rests on the endpoint: HyperBus
has fixed access latency, so the $\tau$ values are constants and the
margin is a true safety margin, whereas a long-tailed endpoint such as an
LPDDR controller makes it a statistical choice.}

\begin{table}[t]
\centering
\caption{WCDT Analysis: Theoretical Models vs. Experimental Configuration for 32-beat Bursts.}
\label{tab:wcdt_eval}
\resizebox{\columnwidth}{!}{
\begin{tabular}{llrrr}
\toprule
\textbf{Arch} & \textbf{Failing Phase / Interval ($i$)} & \textbf{$\tau$ (cyc)} & \textbf{Th (cyc)} & \textbf{WCDT} \\
\midrule
\multirow{7}{*}{PLT}  & AW/AR Handshake ($C_{w0/r0}$)  & 0     & 6     & 6   \\
                      & AW-to-W1 Gap ($C_{w1}$)        & 1     & 9     & 10  \\
                      & First W-beat ($C_{w2}$)        & 2     & 16    & 18  \\
                      & Data Burst Stall ($C_{w3a}$)   & 182   & 43    & 225 \\
                      & Missing B Response ($C_{w4}$)  & 186   & 128   & 314 \\
                      & AR-to-R1 Gap ($C_{r1}$)        & 1     & 65    & 66  \\
                      & Read Burst Stall ($C_{r3a}$)   & 301   & 48    & 349 \\
\midrule
\multirow{3}{*}{CLT}  & AW/AR Handshake                & 0     & 6     & 6   \\
                      & W-Channel Liveness             & 182   & 43    & 225 \\
                      & B-Channel Liveness             & 186   & 128   & 314 \\
                      & R-Channel Liveness             & 301   & 74    & 375 \\
\midrule
\multirow{3}{*}{ILT}  & AW/AR Handshake                & 0     & 6     & 6   \\
                      & Inter-Completion (Write)       & -     & 962   & 962 \\
                      & Inter-Completion (Read)        & -     & 712   & 712 \\
\bottomrule
\end{tabular}
}
\vspace{-6pt}
\end{table}

\begin{figure*}[htbp]
    \centering
    \includegraphics[width=0.95\linewidth]{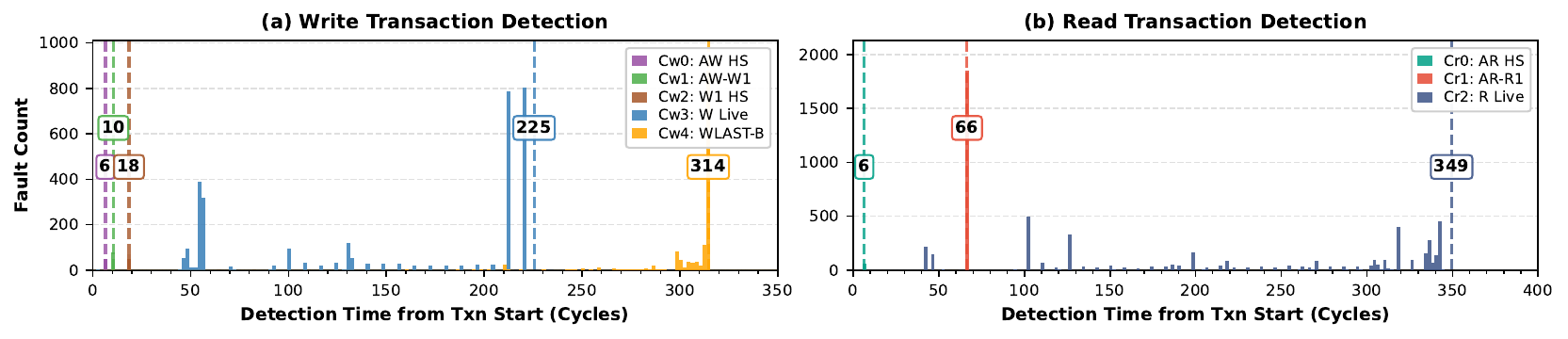}
    \vspace{-5pt}
    \caption{PLT detection latency distributions by transaction phase.}
    \vspace{-15pt}
    \label{fig:plt_phase_timing}
\end{figure*}

\begin{figure*}[htbp]
    \centering
    \includegraphics[width=0.95\linewidth]{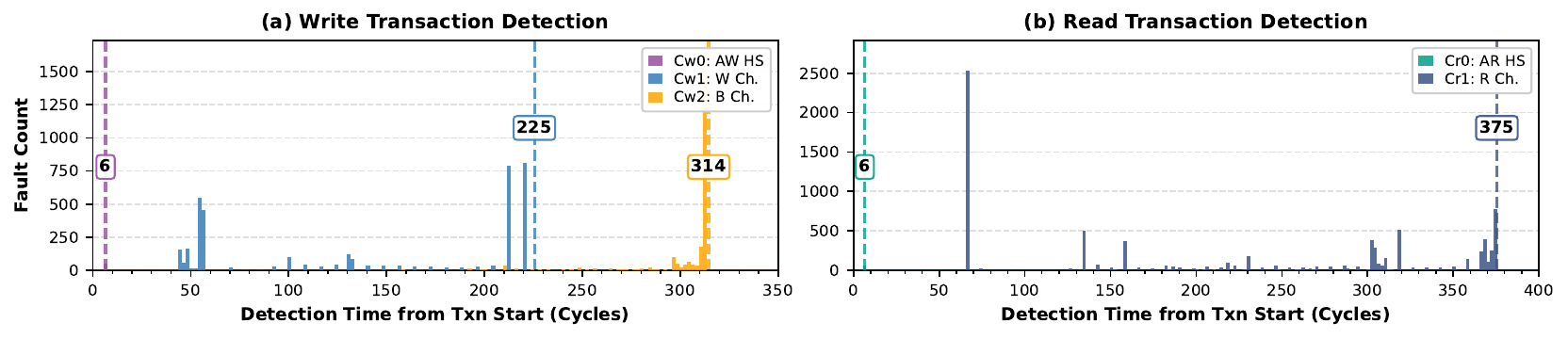}
    \vspace{-5pt}
    \caption{CLT detection latency distributions grouped by AXI channel.}
    \vspace{-15pt}
    \label{fig:clt_phase_timing}
\end{figure*}

\begin{figure*}[htbp]
    \centering
    \includegraphics[width=0.95\linewidth]{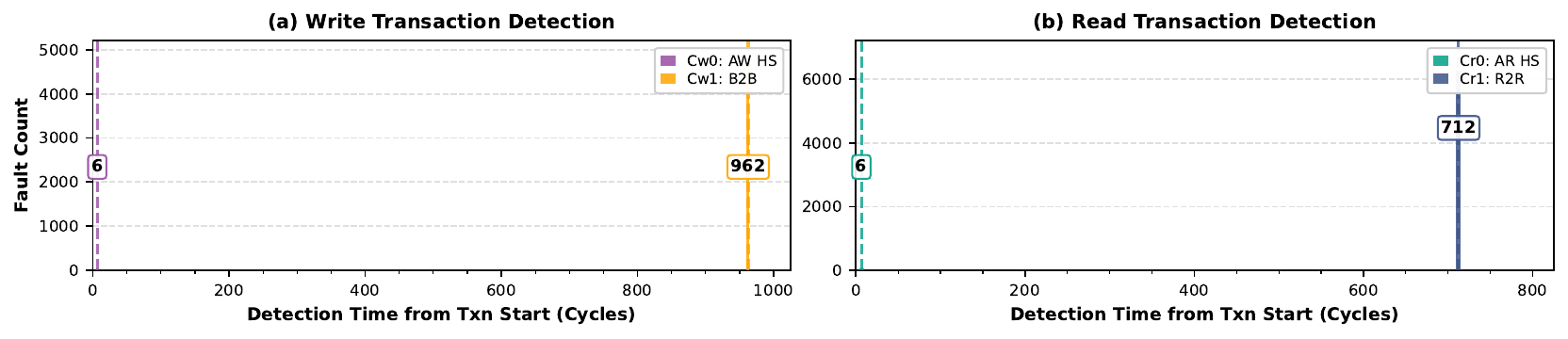}
    \vspace{-5pt}
    \caption{ILT detection latency distributions grouped by transaction ID.}
    \vspace{-15pt}
    \label{fig:ilt_phase_timing}
\end{figure*}

\begin{figure*}[t]
    \centering
    \includegraphics[width=0.8\textwidth]{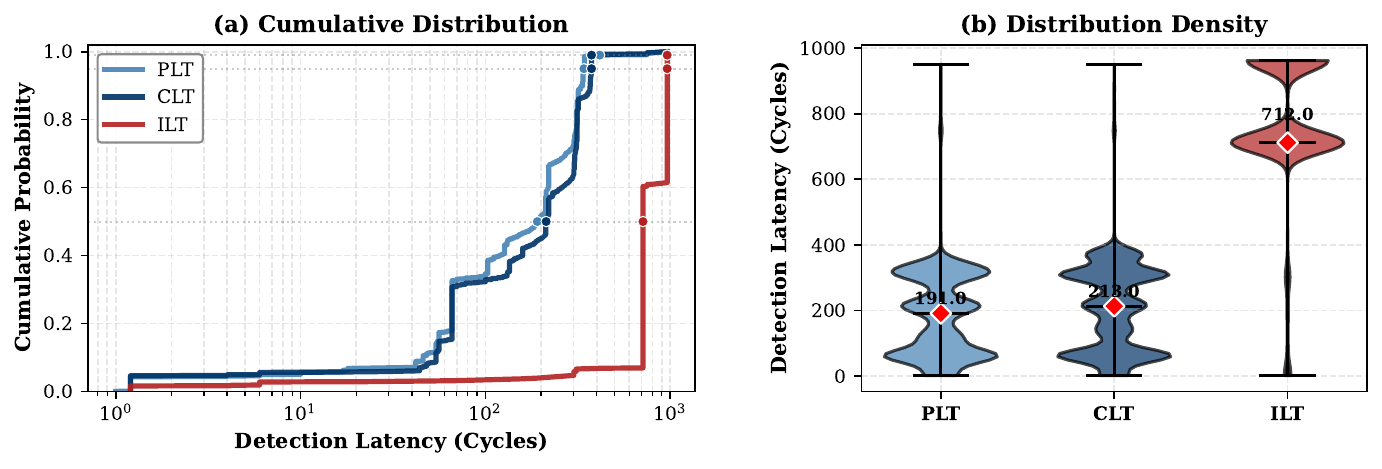}
    \vspace{-5pt}
    \caption{Detection latency comparison across TMU architectures.}
    \label{fig:latency_comparison}
    \vspace{-12pt}
\end{figure*}

\begin{figure}[htbp]
    \centering
    \includegraphics[width=0.38\textwidth, height=0.26\textwidth]{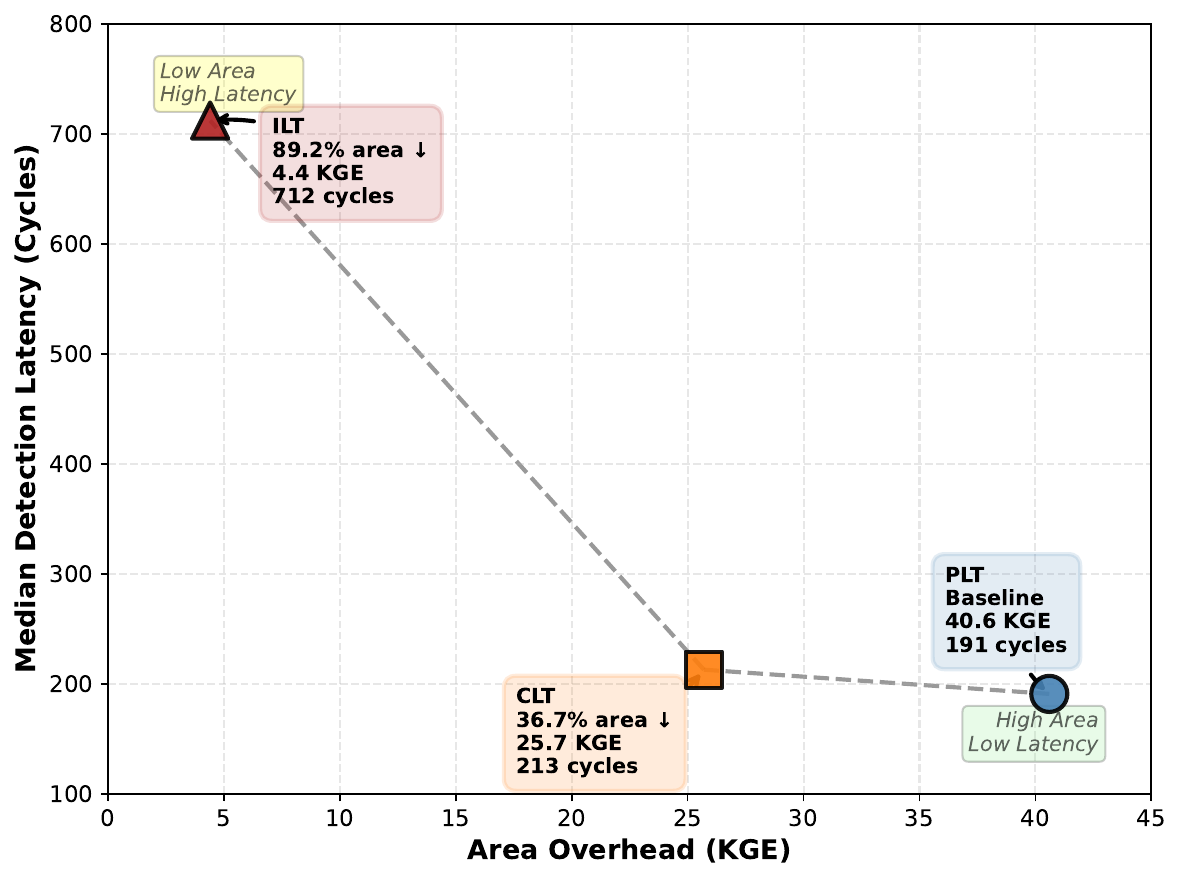}
    \vspace{-5pt}  
    \caption{Area-latency trade-off across TMU architectures at 4×16 configuration.}
    \label{fig:area_latency_tradeoff}
    \vspace{-15pt}
\end{figure} 

Figures~\ref{fig:plt_phase_timing}--\ref{fig:ilt_phase_timing} validate the theoretical WCDT bounds of Table~\ref{tab:wcdt_eval} through fault injection. PLT (Figure~\ref{fig:plt_phase_timing}) produces sharp latency 
concentrations at handshake phases, where detection is deterministic. Liveness monitoring at the data phase exhibits spread below the WCDT bound, as timeouts trigger at any inter-beat interval depending on where the fault disrupts data flow. The spread in \texttt{WLAST}-to-B detections arises because faults perturbing HyperBus backpressure (\texttt{WREADY}) can force \texttt{WLAST} to arrive early, starting the B-response timer ahead of schedule; the fault is nonetheless always caught before the user-configured 314-cycle bound. CLT (Figure~\ref{fig:clt_phase_timing}) subsumes intermediate phases into unified channel boundaries. Early data-phase faults caught by PLT at 10 and 18 cycles shift to the W-channel liveness timeout at 43 cycles, where the per-beat stall is first detected upon liveness threshold expiry. The dominant 225- and 314-cycle worst-case bounds are fully preserved, validating the 36.7\% area reduction at minimal detection latency overhead. ILT (Figure~\ref{fig:ilt_phase_timing}) takes this architectural abstraction to its extreme by stripping away all intra-transaction monitoring. It collapses every intermediate fault into near-impulse distributions at the end-to-end bounds (962 cycles for consecutive writes, 712 cycles for reads). Despite incurring the highest detection latency and sacrificing precise localization, ILT always guarantees a deterministically bounded WCDT sufficient for baseline system safety.

Figure~\ref{fig:latency_comparison} presents detection latency distributions through cumulative distribution function (CDF) (a) and density (b) visualizations. The CDF shows PLT and CLT overlapping closely, with median latencies of 191 and 213 cycles respectively, confirming that channel-level tracking preserves detection speed despite area reduction. ILT's CDF remains near-zero until the 712-cycle read boundary, then rises sharply to 100\% at the 962-cycle write boundary, yielding a median $3.7\times$ higher than PLT. The density violin plot captures this divergence explicitly: PLT and CLT show continuous profiles distributed across all protocol phases, while ILT concentrates bimodally at the two completion-timeout boundaries. This reflects its reliance on end-to-end transaction timeouts alone.

\subsubsection{Area-Latency Design Space}
Figure~\ref{fig:area_latency_tradeoff} maps area overhead against detection latency to visualize the complete TMU design space. PLT minimizes the fault exposure window (191 cycles) at maximum hardware cost, and provides the richest diagnostics: per-phase localization, fault source identification (manager versus subordinate), and detailed latency logging. CLT is the preferred point for most deployments, holding median latency to 213 cycles for a 36.7\% area reduction while retaining channel-level diagnostic visibility. ILT trades a 712-cycle median for an 89.2\% area reduction, suiting tightly area-constrained endpoints where transaction-level deadlock recovery without phase-level diagnostics is sufficient.

\begin{table*}[!t]
\centering
\caption{Comparison of AXI transaction monitoring solutions.}
\label{tab:related_work}
\renewcommand{\arraystretch}{1.15}
\resizebox{\textwidth}{!}{%
\begin{tabular}{|l|c|c|c|c|c|c|c|c|c|c|c|c|}
\hline
\multicolumn{1}{|c|}{\textbf{Reference}} &
\textbf{\begin{tabular}[c]{@{}c@{}}Target\\Prot.\end{tabular}} &
\textbf{\begin{tabular}[c]{@{}c@{}}HW/SW\\Based\end{tabular}} &
\textbf{\begin{tabular}[c]{@{}c@{}}Timing\\Metrics\end{tabular}} &
\textbf{\begin{tabular}[c]{@{}c@{}}Txn.\\Level\end{tabular}} &
\textbf{\begin{tabular}[c]{@{}c@{}}Phase\\Level\end{tabular}} &
\textbf{\begin{tabular}[c]{@{}c@{}}Prot.\\Check\end{tabular}} &
\textbf{\begin{tabular}[c]{@{}c@{}}Perf.\\Metrics\end{tabular}} &
\textbf{\begin{tabular}[c]{@{}c@{}}Fault\\Detect.\end{tabular}} &
\textbf{\begin{tabular}[c]{@{}c@{}}Bounded\\Det.\textsuperscript{a}\end{tabular}} &
\textbf{\begin{tabular}[c]{@{}c@{}}M.O\\Supp.\textsuperscript{b}\end{tabular}} &
\rev{\textbf{\begin{tabular}[c]{@{}c@{}}Area\textsuperscript{c}\end{tabular}}} &
\textbf{\begin{tabular}[c]{@{}c@{}}Area\\Scale\end{tabular}} \\
\hline\hline
\textit{AMD Perf. Mon.}~\cite{b13}                      & AXI     & HW & $\boldsymbol{\checkmark}$ & $\boldsymbol{\checkmark}$ & $\times$ & $\times$ & $\boldsymbol{\checkmark}$ & $\times$ & $\times$ & $\boldsymbol{\checkmark}$ & \rev{n/r} & $\times$ \\ \hline
\textit{Synopsys Smart Mon.}~\cite{b12}                 & AXI     & SW & $\boldsymbol{\checkmark}$ & $\boldsymbol{\checkmark}$ & $\times$ & $\times$ & $\boldsymbol{\checkmark}$ & $\times$ & $\times$ & $\times$ & \rev{n/r} & $\times$ \\ \hline
Ravi et al.\ (Bus Mon.)~\cite{b_ravi}                   & AXI     & HW & $\boldsymbol{\checkmark}$ & $\boldsymbol{\checkmark}$ & $\times$ & $\times$ & $\boldsymbol{\checkmark}$ & $\times$ & $\times$ & $\times$ & \rev{n/r} & $\times$ \\ \hline
Kyung et al.\ (PMU)~\cite{b5}                           & AXI     & HW & $\boldsymbol{\checkmark}$ & $\boldsymbol{\checkmark}$ & $\times$ & $\times$ & $\boldsymbol{\checkmark}$ & $\times$ & $\times$ & $\times$ & \rev{n/r} & $\times$ \\ \hline
Tan et al.\ (Perf. Eval.)~\cite{b7}                     & AXI     & SW & $\boldsymbol{\checkmark}$ & $\boldsymbol{\checkmark}$ & $\times$ & $\times$ & $\boldsymbol{\checkmark}$ & $\times$ & $\times$ & $\times$ & \rev{n/r} & $\times$ \\ \hline
Edelman et al.\ (Trans. Mon.)~\cite{b8}                 & AXI     & SW & $\times$ & $\boldsymbol{\checkmark}$ & $\boldsymbol{\checkmark}$ & $\times$ & $\times$ & $\times$ & $\times$ & $\times$ & \rev{n/r} & $\times$ \\ \hline\hline
\textit{AMD Prot. Checker}~\cite{amd_pc}                & AXI     & HW & $\times$ & $\boldsymbol{\checkmark}$ & $\times$ & $\boldsymbol{\checkmark}$ & $\times$ & $\times$ & $\times$ & $\boldsymbol{\checkmark}$ & \rev{n/r} & $\times$ \\ \hline
Chen et al.\ (AXIChecker)~\cite{b6}                     & AXI     & HW & $\times$ & $\boldsymbol{\checkmark}$ & $\times$ & $\boldsymbol{\checkmark}$ & $\times$ & $\times$ & $\times$ & $\times$ & \rev{70.7\,kGE\textsuperscript{d}} & $\times$ \\ \hline
Lee et al.\ (RecoSoC Mon.)~\cite{b4}                    & AXI     & HW & $\boldsymbol{\checkmark}$ & $\boldsymbol{\checkmark}$ & $\times$ & $\boldsymbol{\checkmark}$ & $\boldsymbol{\checkmark}$ & $\times$ & $\times$ & $\times$ & \rev{n/r} & $\times$ \\ \hline\hline
Lazaro et al.\ (Firewall)~\cite{b2}                     & AXI-Lite     & HW & $\times$ & $\boldsymbol{\checkmark}$ & $\times$ & $\times$ & $\times$ & $\times$ & $\times$ & $\times$ & \rev{0.13\,kLUT\textsuperscript{e}} & $\times$ \\ \hline
Zonta et al.\ (XRAY)~\cite{xray}                        & AXI     & SW & $\times$ & $\boldsymbol{\checkmark}$ & $\circ$ & $\boldsymbol{\checkmark}$ & $\times$ & $\boldsymbol{\checkmark}$ & $\times$ & $\times$ & \rev{0.40\,kLUT\textsuperscript{f}} & $\times$ \\ \hline
Foudhaili et al.\ (IMS)~\cite{ims}                      & AXI     & HW & $\times$ & $\checkmark$ & $\times$ & $\boldsymbol{\checkmark}$ & $\times$ & $\boldsymbol{\checkmark}$ & $\times$ & $\circ$ & \rev{20.8\,kLUT\textsuperscript{g}} & $\times$ \\ \hline\hline
\textit{Xilinx AXI Timeout}~\cite{b_xilinx_to}         & AXI     & HW & $\boldsymbol{\checkmark}$ & $\boldsymbol{\checkmark}$ & $\times$ & $\times$ & $\times$ & $\boldsymbol{\checkmark}$ & $\times$ & $\times$ & \rev{n/r} & $\times$ \\ \hline
\textit{ARM SP805 Watchdog}~\cite{b11}                  & APB     & HW & $\boldsymbol{\checkmark}$ & $\times$ & $\times$ & $\times$ & $\times$ & $\boldsymbol{\checkmark}$ & $\times$ & $\times$ & \rev{n/r} & $\times$ \\ \hline
Cabo et al.\ (SafeSU)~\cite{b15}                        & AHB/AXI & HW & $\boldsymbol{\checkmark}$ & $\boldsymbol{\checkmark}$ & $\times$ & $\times$ & $\boldsymbol{\checkmark}$ & $\boldsymbol{\checkmark}$ & $\times$ & $\times$ & \rev{5.0\,kLUT\textsuperscript{h}} & $\times$ \\ \hline
Benz et al.\ (AXI-REALM)~\cite{b_realm}                 & AXI     & HW & $\boldsymbol{\checkmark}$ & $\boldsymbol{\checkmark}$ & $\boldsymbol{\checkmark}$ & $\boldsymbol{\checkmark}$ & $\boldsymbol{\checkmark}$ & $\boldsymbol{\checkmark}$ & $\times$ & $\boldsymbol{\checkmark}$  & \rev{50.0\,kGE\textsuperscript{i}} & $\boldsymbol{\checkmark}$ \\ \hline\hline
\rowcolor{rowgray}\textbf{This work: PLT}       & AXI     & HW & $\boldsymbol{\checkmark}$ & $\boldsymbol{\checkmark}$ & $\boldsymbol{\checkmark}$ & $\boldsymbol{\checkmark}$ & $\boldsymbol{\checkmark}$ & $\boldsymbol{\checkmark}$ & $\boldsymbol{\checkmark}$ & $\boldsymbol{\checkmark}$ & \rev{\textbf{40.6\,kGE\textsuperscript{j}}} & $\boldsymbol{\checkmark}$ \\ \hline
\rowcolor{rowgray}\textbf{This work: CLT}       & AXI     & HW & $\boldsymbol{\checkmark}$ & $\boldsymbol{\checkmark}$ & $\boldsymbol{\checkmark}$ & $\boldsymbol{\checkmark}$ & $\boldsymbol{\checkmark}$ & $\boldsymbol{\checkmark}$ & $\boldsymbol{\checkmark}$ & $\boldsymbol{\checkmark}$ & \rev{\textbf{25.7\,kGE\textsuperscript{j}}} & $\boldsymbol{\checkmark}$ \\ \hline
\rowcolor{rowgray}\textbf{This work: ILT}       & AXI     & HW & $\boldsymbol{\checkmark}$ & $\boldsymbol{\checkmark}$ & $\times$ & \rev{$\circ$\textsuperscript{k}} & $\boldsymbol{\checkmark}$ & $\boldsymbol{\checkmark}$ & $\boldsymbol{\checkmark}$ & $\boldsymbol{\checkmark}$ & \rev{\textbf{4.4\,kGE\textsuperscript{j}}} & $\boldsymbol{\checkmark}$ \\ \hline
\end{tabular}%
}
\begin{minipage}{\linewidth}
\vspace{2pt}
\footnotesize
$\boldsymbol{\checkmark}$ = supported; \rev{$\circ$ = partially supported;} $\times$ = not supported. \textit{Italic} = commercial tools.
\textsuperscript{a}~Detection within a formally derived worst-case bound validated by fault injection.
\textsuperscript{b}~Multiple outstanding transaction support.
\rev{\textsuperscript{c}~As published, not normalized; n/r = not reported or software-based.
\textsuperscript{d}~TSMC 180\,nm.
\textsuperscript{e}~Zynq.
\textsuperscript{f}~VCU118, full-mode checker.
\textsuperscript{g}~ZCU104.
\textsuperscript{h}~KCU105.
\textsuperscript{i}~GF12, erealm unit.
\textsuperscript{j}~GF12, $4\times16$.
\textsuperscript{k}~Response-ID checking only; no burst-length validation.}
\end{minipage}
\vspace{-12pt}
\end{table*}
\vspace{-5pt}
\section{Related Work}
\label{sec:related}

Existing transaction monitoring solutions address four distinct challenges: performance characterization, protocol compliance, access control and security, and runtime fault detection. Each category addresses only a subset of the requirements imposed by safety-critical mixed-criticality SoCs, as summarized in Table~\ref{tab:related_work}.

\textit{Performance Monitoring.}
These solutions measure AXI traffic for system tuning but do not detect faults. Ravi~\cite{b_ravi} implemented a hardware bus monitor that records transfer counts, sizes, and separate read and write latency counts for post-silicon analysis. Kyung et al.~\cite{b5} developed an AXI monitoring unit that gathers bandwidth statistics across user-defined address ranges. Tan et al.~\cite{b7} proposed a simulation-based UVM approach measuring AXI bandwidth and latency. Commercial tools include the AMD AXI Performance Monitor~\cite{b13}, a synthesizable hardware IP, and Synopsys Smart Monitor~\cite{b12}, a DPI-based pre-silicon tool; both target performance characterization without runtime fault detection. Edelman~\cite{b8} created a UVM transaction-monitoring framework, likewise pre-silicon only.

\textit{Protocol Checking.}
These verify AXI protocol rules but ignore timing and liveness. Chen et al.~\cite{b6} developed AXIChecker, a synthesizable checker enforcing 44 AXI compliance rules via on-chip assertions. The AMD AXI Protocol Checker~\cite{amd_pc} provides assertion logic for design-time verification of channel handshakes. Lee and Huang~\cite{b4} proposed a reconfigurable monitor combining protocol checking with latency measurement, but it fails to detect critical AXI4 ordering violations such as ID mismatches or out-of-order completions within the same ID and lacks fault recovery mechanisms.

\textit{Access Control and Security.}
These filter traffic or detect intentional attacks but ignore timing faults. Lazaro et al.~\cite{b2} implemented an AXI firewall that blocks traffic violating predefined address ranges and bandwidth limits. Zonta et al.~\cite{xray} created XRAY, an offline tool that identifies architectural vulnerabilities by checking 13 security properties across seven AXI interconnects. Foudhaili et al.~\cite{ims} developed IMS, an ML-based monitor using a quantized neural network to detect denial-of-service attacks. IMS complements the TMU: it classifies transaction headers to detect intentional security breaches, while the TMU detects unintentional liveness violations throughout the entire transaction lifecycle. Pagani et al.~\cite{pagani2019}, Restuccia et al.~\cite{restuccia2019}, and AXI HyperConnect~\cite{hyperconnect} enforce bandwidth reservation, arbiter fairness, and hypervisor-level isolation respectively for AXI accelerators on FPGAs. These works address resource access control but do not detect protocol-level faults or restore interconnect liveness upon subordinate failure.

\textit{Timeout and Fault Detection.}
These detect stalls but lack fine-grained diagnostics. The Xilinx AXI Timeout IP~\cite{b_xilinx_to} measures time between address handshake and response arrival but lacks visibility into data phases or support for multiple outstanding transactions. The ARM SP805 Watchdog~\cite{b11} provides coarse system-level protection over APB but lacks AXI transaction awareness. Cabo et al.~\cite{b15} developed SafeSU, which monitors per-master bus contention to enforce interference quotas between managers. While effective for WCET predictability and contention mitigation, SafeSU does not perform per-phase protocol checks or AXI4-specific burst validation. Benz et al.~\cite{b_realm} introduced AXI-REALM, which monitors traffic at interconnect ingress and egress to enforce bandwidth budgets and traffic regulation; its egress monitoring builds on the preliminary TMU of~\cite{b30}, which this work substantially extends with formal WCDT analysis and a three-tier architecture.

\rev{Table~\ref{tab:related_work} reports published area where available.
Among the ASIC entries, AXIChecker~\cite{b6} occupies 70.7\,kGE in
180\,nm, enforcing protocol rules without timeout supervision,
multiple-outstanding support, or recovery, whereas PLT provides all three
in 40.6\,kGE; the node difference precludes a direct comparison, but
AXIChecker's functional scope is a strict subset of PLT's. ILT provides
bounded detection and interconnect recovery in 4.4\,kGE, an order of
magnitude below the prior ASIC entries, while the coarse timeout
mechanisms in the table provide neither.}

\textit{Prior Work.}
A preliminary work~\cite{b30} introduced the TMU concept with
Tiny-Counter and Full-Counter variants. This current work extends that
design with the three-tier granularity framework derived from AXI4
ordering rules, and with WCDT analysis validated across 1.2 million
fault-injection scenarios. PLT achieves a 62.7\% area reduction at the
$4\times16$ configuration (108.8\,kGE to 40.6\,kGE), CLT reduces area by
40.3\% over the Tiny-Counter while providing strictly finer channel-level
localization, and ILT introduces sub-linear
$\mathcal{O}(N_{ID})$ scaling with no counterpart in the preliminary
design.
\vspace{-5pt}
\section{Conclusion}
\label{sec:conclusion}
This paper presented a TMU framework for AXI4-based safety-critical SoCs, spanning the monitoring granularity and area trade-off through three variants: PLT for per-phase fault localization, CLT consolidating phase monitors into channel-level watchers at a 36.7\% area reduction, and ILT eliminating per-transaction state for sub-linear scaling at an 89.2\%reduction. 
Fault injection on a RISC-V SoC confirmed that no fault manifesting as an AXI4 protocol or liveness violation escaped detection, with latencies bounded by the analytically derived Worst-Case Detection Time. Together they define a structured design space for aligning monitoring overhead with subordinate criticality and safety certification requirements.

\bibliographystyle{IEEEtran}

\end{document}